\documentclass[aps,nofootinbib,twocolumn,superscriptaddress]{revtex4}
\usepackage{mathtools,braket,slashed,bm}
\usepackage{booktabs}
\usepackage{setspace}
\usepackage{siunitx}
\usepackage{multirow}
\usepackage{color}
\usepackage{ulem} 
\usepackage{tensor}

\usepackage{mathrsfs}

\usepackage[utf8]{inputenc}

\renewcommand\sout{\bgroup \color{red} \ULdepth=-.5ex \ULset}

\begin{document}

\title{Analysis of nuclear structure in a converging expansion scheme}

\author{Hana \surname{Gil}}
\email{gil@knu.ac.kr}
\affiliation{Department of Physics, Kyungpook National University, Daegu 41566, Korea}

\author{Young-Min \surname{Kim}}
\email{ymkim715@gmail.com}
\affiliation{School of Natural Science, Ulsan National Institute of Science and Technology, Ulsan 44919, Korea}

\author{Chang Ho \surname{Hyun}}
\email{hch@daegu.ac.kr}
\affiliation{Department of Physics Education, Daegu University, Gyeongsan 38453, Korea}

\author{Panagiota \surname{Papakonstantinou}}
\email{ppapakon@ibs.re.kr}
\affiliation{Rare Isotope Science Project, Institute for Basic Science, Daejeon 34047, Korea}

\author{Yongseok Oh}
\email{yohphy@knu.ac.kr}
\affiliation{Department of Physics, Kyungpook National University, Daegu 41566, Korea}
\affiliation{Asia Pacific Center for Theoretical Physics, Pohang, Gyeongbuk 37673, Korea}

\date{\today}

\begin{abstract}

\textbf{Background:} In the framework of the newly developed generalized energy density functional (EDF) called KIDS, 
the nuclear equation of state (EoS) is expressed as an expansion in powers of the Fermi momentum or the cubic root of the 
density ($\rho^{1/3}$). 
Although an optimal number of converging terms was obtained in specific cases of fits to empirical data and pseudodata, 
the degree of convergence remains to be examined not only for homogeneous matter but also for finite nuclei.
Furthermore, even for homogeneous matter, the convergence should be investigated with widely adopted various EoS properties 
at saturation.  

\textbf{Purpose:} The first goal is to validate the minimal and optimal number of EoS parameters required for 
the description of homogeneous nuclear matter over a wide range of densities relevant for astrophysical applications. 
The major goal is to examine the validity of the adopted expansion scheme for an accurate description of finite nuclei.   

\textbf{Method:} We vary the values of the high-order density derivatives of the nuclear EoS, such as the skewness of the energy 
of symmetric nuclear matter and the kurtosis of the symmetry energy, at saturation and examine the relative importance of each term in 
$\rho^{1/3}$ expansion for homogeneous matter. 
For given sets of EoS parameters determined in this way, we define equivalent Skyrme-type functionals and examine the convergence 
in the description of finite nuclei focusing on the masses and charge radii of closed-shell nuclei. 

\textbf{Results:} The EoS of symmetric nuclear matter is found to be efficiently parameterized with only 3 parameters 
and the symmetry energy (or the energy of pure neutron matter) with 4 parameters when the EoS is expanded in the power series 
of the Fermi momentum. 
Higher-order EoS parameters do not produce any improvement, in practice, in the description of nuclear ground-state energies 
and charge radii, which means that they cannot be constrained by bulk properties of nuclei. 

\textbf{Conclusions:} The minimal nuclear EDF obtained in the present work is found to reasonably describe the properties of closed-shell
nuclei and the mass-radius relation of neutron stars.
Attempts at refining the nuclear EDF beyond the minimal formula must focus on parameters which are not active 
(or strongly active) in unpolarized homogeneous matter, for example, effective tensor terms and time-odd terms.


\end{abstract}

\maketitle

\section{Introduction}

In a series of publication~\cite{PPLH16,GOHP17,GPHPO16,GPHO18}, we have proposed and developed a strategy to model 
nuclear systems based on a converging power expansion combined with energy density functional (EDF) theory. 
Beginning with homogeneous matter~\cite{PPLH16}, we formulated the energy per particle, which represents the equation of state (EoS),
as an expansion in powers of the Fermi momentum or equivalently in powers of the cubic root of the density, as $k_F\propto \rho^{1/3}$. 
This choice is rooted both in quantum many-body theory and effective field theory.
We confirmed \textit{a posteriori\/} the quick convergence of the expansion by fitting the parameters to pseudodata from microscopic calculations.
Based on a statistical analysis of the fits, a robust parameter set was chosen as a baseline for further explorations, 
comprising three terms for isospin-symmetric nuclear matter (SNM) and four for pure neutron matter (PNM). 
The naturalness of the expansion was confirmed and extrapolations to extreme density regimes, were found to be satisfactory~\cite{GPHO18}. 
In particular, the extrapolated results agreed with \textit{ab initio\/} calculations for dilute neutron matter, a regime to which the model had not been fitted at all, 
and reproduced a realistic mass-radius relation of neutron stars, which represents a dense regime.

In subsequent works reported in Refs.~\cite{GOHP17, GPHPO16, GPHO18}, the KIDS EoS was transposed to a Skyrme functional with extended density-dependent couplings, 
which we call a KIDS EDF,  
to study nuclear ground-state properties, thereby relying on the Kohn-Sham scheme~\cite{KS65,BHR03}.%
With the baseline EoS from Ref.~\cite{PPLH16} and only six input data, namely the ground-state energies and charge radii of three nuclei, 
it was possible to obtain a successful description of the bulk properties of closed-(sub)shell nuclei over a wide range of atomic number, 
say from \nuclide[16]{O} to \nuclide[218]{U}~\cite{GOHP17,GPHPO16,GPHO18}.%
\footnote{Because only closed-(sub)shell nuclei were considered, we do not include pairing interactions in the present work.}
Furthermore, the results were found to be practically independent of the assumption on the in-medium effective mass~\cite{GPHO18}, 
which means that the latter cannot be efficiently constrained by the bulk static properties of nuclei.   
The corresponding parameters then remain to be determined via dynamic properties of nuclei.
The above results showed that with a well-defined nuclear EoS Ansatz, the convenient Skyrme formalism, and simple rules 
for fitting, it would be possible to find a \textit{unified and phenomenological\/} nuclear model describing nuclear matter and nuclei
with the same parameter set, i.e., the same EoS.

Before developing more sophisticated models to describe various types of nuclei along this approach, 
we address the convergence issue in the description of closed-(sub)shell nuclei at the present stage. 
Throughout the previous publications~~\cite{PPLH16,GOHP17,GPHPO16,GPHO18}
we have shown that the expansion of the EoS as a power series of the Fermi momentum exhibits excellent convergence well above the
saturation density~\cite{GOHP17}.
However, careful analyses lead to the observation that the degree of convergence depends on isospin and, as a result, higher order 
contributions are more important in PNM than in SNM.
In fact, in Ref.~\cite{PPLH16}, it is shown that, while three terms are sufficient for describing SNM in a fast-converging hierarchy,
at least four terms are needed to have such behavior for PNM. 
The origin of this difference is certainly of theoretical interest and requires sophisticated investigations on nuclear dynamics. 
Although we will not address here the issue on its fundamental origins, it would be important and meaningful to examine the convergence 
in the description of finite nuclei.    
This is the major motivation of the present article and the purpose of the this work is to examine the convergence 
of the power series expansion in the Fermi momentum for the description of finite nuclei.

The nuclear EoS is often represented in terms of parameters defined at the saturation point such as the saturation density $\rho_0^{}$, 
the binding energy per particle at saturation $E_0^{}$, the symmetry energy at saturation $J$, the slope parameter $L$,
and the compression modulus $K_0$. 
These parameters were used to constrain the nuclear EoS in our previous publications~\cite{PPLH16,GOHP17,GPHPO16,GPHO18}.
However, the role of the parameters that are related to higher derivatives of the EoS with respect to density remains to be explored.
These ``EoS parameters’’ can be readily expressed analytically in terms of the KIDS expansion coefficients. 
The question of how many KIDS parameters are needed for an efficient description of nuclear systems can be rephrased 
as how many high-order derivatives of the SNM energy and of the symmetry energy are needed. 
In other words, we also need to examine how many EoS parameters are necessary for an efficient and 
well-converged description of PNM and nuclear ground states.
Furthermore, since higher-order terms in the power series expansion control the behavior of the nuclear EoS at higher densities, 
higher-order EoS parameters such as the skewness and kurtosis would help in constraining the EoS at higher densities and 
examining the convergence of the EoS.

Motivated by the above issues, in the present work we address the following questions.
In Refs.~\cite{PPLH16,GOHP17, GPHPO16, GPHO18}, we successfully parameterized the EoS of SNM and PNM by three  and four
EoS parameters in the considered range of densities. 
Then it is natural to seek how far the constructed EoS can be applied as a function of density. 
This is related to the role of higher order EoS parameters and we explore the sensitivity of our EoS to higher-order EoS parameters.
Once their role is identified for homogeneous nuclear matter, we investigate the role of higher-order EoS parameters in the description of
finite nuclei.
To this end, we obtain results for various values of the skewness of the SNM EoS and the kurtosis of the symmetry energy at the saturation point
to confirm that such higher order terms hardly play any role.  
The corollary is that the skewness of the SNM EoS and the symmetry-energy kurtosis cannot be practically constrained 
by the static properties of nuclei such as masses and radii.

This paper is organized as follows. 
In Sec.~\ref{kids}, we briefly review the formalism of the KIDS EDF and the corresponding Skyrme potentials will be developed. 
Section~\ref{symmetric} is devoted to the exploration of the uncertainty in the fourth-order term in SNM and the role of the skewness
of the SNM EoS is examined.
The mass-radius relations of neutron stars are also computed within the models of the present approach.
Then, in Sec.~\ref{asymmetric}, we increase the number of terms in the asymmetric part of EDF to investigate the
convergence behavior of the model with respect to the kurtosis of the nuclear symmetry energy.
In Sec.~\ref{discussion}, we discuss the results in the context of current efforts to extend the nuclear EDF, 
in particular, in the form of extended Skyrme functionals with rich momentum dependence and tensor forces.  
Finally, we summarize and conclude in Sec.~\ref{summary}.

\section{KIDS EDF: equation of state and corresponding Skyrme functionals\label{kids}}

\subsection{KIDS equation of state} 

In the KIDS model for nuclear EDF, the energy per particle in homogeneous nuclear matter is expanded in powers of the Fermi momentum
$k_F$ or equivalently the cubic root of the baryon density $\rho$.
Thus the nuclear EDF in this approach is written as
\begin{eqnarray}
\mathscr{E}(\rho , \delta ) = \mathscr{T}(\rho,\delta ) + \sum_{i=0}^{N-1} c_i (\delta) \rho^{1+i/3}
\label{eq1},
\end{eqnarray}
where $\mathscr{T}$ is the free Fermi-gas kinetic energy and the potential energy is expanded up to $N$ terms, 
namely, up to the order of $\rho^{(N+2)/3}$ starting from the $\rho$ term.
The isospin asymmetry $\delta$ is defined as $\delta = (\rho_n^{} - \rho_p^{})/\rho$, where $\rho_n^{}$ and $\rho_p^{}$ 
are neutron and proton densities, respectively, which give the total nucleon density $\rho = \rho_n^{}+\rho_p^{}$.
Model parameters $c_i^{} (\delta)$ could be expanded in even powers of isospin asymmetry $\delta$.
For the purpose of the present work, we adopt the usual quadratic approximation for the isospin-asymmetry dependence of 
$c_i^{} (\delta)$ by writing
\begin{eqnarray}
c_i^{} (\delta)  = \alpha_i^{} + \beta_i^{} \delta^2,
\label{eq:cdelta}
\end{eqnarray}
which leads to $\alpha_i^{} = c_i^{} (0)$ and $\beta_i^{} = c_i^{} (1) - c_i^{} (0)$.

The expansion parameters $c_i^{}(\delta)$ can be constrained once the empirical properties of nuclear matter, i.e., EoS parameters, are known.
Phenomenologically, these parameters are defined at nuclear saturation density by the series expansion of the SNM energy $\mathscr{E}(\rho,0)$ 
and nuclear symmetry energy, which can be defined and expressed as 
\begin{equation} 
\mathcal{S}(\rho ) =  \left. \frac12 \frac{\partial^2}{\partial \delta^2} \mathscr{E}(\rho , \delta ) \right |_{\delta = 0}
= \mathscr{T}_{\mathrm{sym}}(\rho ) + \sum_{i=0}^{N-1} \beta_i \rho^{1+i/3}
\label{Eq:SymEn}, 
\end{equation} 
where the contribution from the free kinetic energy reads
\begin{eqnarray}
\mathscr{T}_{\mathrm{sym}}(\rho) = \frac{\hbar^2}{6\, m} \left( \frac{3 \pi^2}{2} \right)^{2/3} \rho^{2/3}.
\end{eqnarray} 
Then EoS parameters of interest are defined through~\cite{DLSD12}
\begin{eqnarray}
\mathscr{E}(\rho,0) &=& E_0^{} + \frac12 K_0 x^2 + \frac16 Q_0 x^3 + O(x^4), \nonumber \\
\mathcal{S}(\rho) &=& J + Lx + \frac12 K_{\mathrm{sym}} x^2 + \frac16 Q_{\mathrm{sym}} x^3 + \frac{1}{24} R_{\mathrm{sym}} x^4
\nonumber \\ && \mbox{}
 + O(x^5),
\end{eqnarray}
where $x = (\rho - \rho_0^{})/(3 \rho_0^{})$.

Therefore, the SNM energy is characterized by the saturation density $\rho_0^{}$, the energy per particle at saturation $E_0^{}$, 
the compression modulus $K_0$, and the skewness coefficient $Q_0$ defined as
\begin{eqnarray}
K_0 &\equiv& \left. 9 \rho^2_0 \, \frac{d^2}{d\rho^2} \frac{\mathscr{E}(\rho ,0)}{\rho} \right|_{\rho=\rho_0^{}},
\nonumber \\
Q_0 &\equiv& \left. 27 \rho_0^3 \, \frac{d^3}{d\rho^3} \mathscr{E}(\rho ,0) \right|_{\rho=\rho_0^{}}.
\end{eqnarray}
On the other hand, the nuclear symmetry energy is customarily characterized at the saturation point by its value 
$J=\mathcal{S}(\rho_0^{})$, the slope $L$, and the curvature $K_{\mathrm{sym}}$ defined as
\begin{eqnarray}
L &\equiv&  \left. 3 \rho_0 \, \frac{d}{d\rho} \mathcal{S}(\rho ) \right|_{\rho=\rho_0^{}},
\nonumber \\
K_{\mathrm{sym}} &\equiv& \left. 9 \rho^2_0 \, \frac{d^2}{d\rho^2} \frac{\mathcal{S}(\rho)}{\rho} \right|_{\rho=\rho_0^{}}.
\end{eqnarray}
In addition, we consider the skewness $Q_{\mathrm{sym}}$ and the kurtosis $R_{\mathrm{sym}}$, 
defined via the 3rd and 4th derivatives, respectively, as
\begin{eqnarray}
 Q_{\mathrm{sym}} &\equiv& \left. 27 \rho_0^3 \, \frac{d^3}{d\rho^3} \mathcal{S}(\rho ) \right|_{\rho=\rho_0^{}},
 \nonumber \\
 R_{\mathrm{sym}} &\equiv& \left. 81 \rho_0^4 \, \frac{d^4}{d\rho^4} \mathcal{S}(\rho ) \right|_{\rho=\rho_0^{}}.
\end{eqnarray} 
These EoS parameters will be discussed in the parameterization of the KIDS model.

All the above quantities are readily obtained analytically with the help of expressions of Eqs.~(\ref{eq1})--(\ref{Eq:SymEn}). 
Explicitly, we have
\begin{eqnarray} \label{eq:eosB} 
K_0 &=& -2\mathscr{T}(\rho_0^{},0) + \sum_{i=0}^{N-1} i(i+3) \alpha_i \rho_0^{1+i/3},  \\
Q_0 &=& +8\mathscr{T}(\rho_0^{},0) + \sum_{i=0}^{N-1} i(i+3)(i-3) \alpha_i \rho_0^{1+i/3}, \\
K_{\mathrm{sym}} &=& -2\mathscr{T}_{\mathrm{sym}} (\rho_0^{}) + \sum_{i=0}^{N-1} i(i+3) \beta_i \rho_0^{1+i/3}, \\
Q_{\mathrm{sym}} &=& +8\mathscr{T}_{\mathrm{sym}} (\rho_0^{}) + \sum_{i=0}^{N-1} i(i+3)(i-3) \beta_i \rho_0^{1+i/3},\\ 
R_{\mathrm{sym}} &=& -56\mathscr{T}_{\mathrm{sym}} (\rho_0^{}) + \sum_{i=0}^{N-1} i(i+3)(i-3)(i-6) \beta_i \rho_0^{1+i/3}. \nonumber \\
\label{eq:eosE} 
\end{eqnarray} 
These relations connect the values of the EoS parameter to our model parameters $\alpha_i$ and $\beta_i$.
Once the values of EoS parameters are known, our approach allows us to find the nuclear EoS to the desired order in density. 
However, most of the above EoS parameters are not known to a satisfactory accuracy and ranges of their values are to be explored.

In Ref.~\cite{PPLH16}, we determined the baseline KIDS parameter set labeled `KIDS-ad2' in the following way. 
We began by fitting many possible combinations (of varying order $N$) of KIDS parameters $\alpha_i^{}$ and $\beta_i^{}$ to the 
Akmal-Pandharipande-Ravenhall (APR) EoS~\cite{APR98}. 
Having concluded that the three lowest-order terms are sufficient for the description of SNM, we set $\alpha_3^{}=0$, 
and determined $\alpha_{0,1,2}^{}$ by widely adopted empirical properties at saturation, namely, $\rho_0^{}=0.16$~$\mbox{fm}^{-3}$, 
$E_0^{} = -16$~MeV, and $K_0=240$~MeV.%
(These values are also consistent with the APR EoS.)
This model is then found to give the skewness coefficient $Q_0 \approx -373$~MeV. 
The coefficients $c_i(1)$, or equivalently $\beta_i$, were also fitted to the APR EoS for PNM. 
In this case, we found that at least four terms had to be retained in the KIDS EDF in order to reproduce the APR EoS for PNM. 
The resulting EDF gives $J=32.8$~MeV, $L=49.3$~MeV, $K_{\rm sym} = -156$~MeV, $Q_{\rm sym}=583$ MeV, and
$R_{\rm sym} = -2470$~MeV.

The KIDS-ad2 EoS determined in this way was subsequently transposed into a zero-range, density-dependent effective interaction for nuclei 
and applied successfully to Hartree-Fock calculations of nuclear ground states of closed-shell nuclei~\cite{GPHPO16,GPHO18}, 
providing satisfactory results, on a par with generalized Skyrme-type functionals. 
The question to be addressed at the present work is to examine whether superior results can be obtained with higher-order terms.

\subsection{Corresponding Skyrme functionals 
\label{Sec:Skyrme}}

In this subsection, we review a simple procedure for applying a given KIDS EoS to the description of finite nuclei, which will be employed in the present work. 
The Fermi momentum expansion of EDF in Eq.~(\ref{eq1}) leads to a convenient Skyrme-type effective interaction~\cite{GPHO18} in the form of
\begin{eqnarray}
v_{ij}^{}
&=& (t_0^{} + y_0^{} P_\sigma) \delta(\mathbf{r}_i-\mathbf{r}_j)  
\nonumber \\ && \mbox{} 
  +\frac{1}{2} (t_1^{} + y_1^{} P_\sigma)[\delta(\mathbf{r}_i^{} - \mathbf{r}_j^{})\mathbf{k}^2 
  + {\mathbf{k}^\prime}^2\delta(\mathbf{r}_i^{} - \mathbf{r}_j^{})] 
\nonumber \\ && \mbox{} 
  + (t_2^{} + y_2^{} P_\sigma)\mathbf{k}^\prime \cdot \delta(\mathbf{r}_i^{} - \mathbf{r}_j^{} )\mathbf{k} 
\nonumber \\ && \mbox{} 
  + \frac{1}{6} \sum_{n=1}^{N-1}(t_{3n}^{} + y_{3n}^{} P_\sigma)\rho^{n/3}\delta(\mathbf{r}_i^{} - \mathbf{r}_j^{}) 
\nonumber \\ && \mbox{} 
+ iW_0\,\mathbf{k}^\prime \times \delta(\mathbf{r}_i^{} - \mathbf{r}_j^{})\,
\mathbf{k}  \cdot (\boldsymbol{\sigma}_i^{} - \boldsymbol{\sigma}_j^{}),	
\label{eq:skyrme}	
\end{eqnarray}
where $\mathbf{k} = (\bm{\nabla}_i - \bm{\nabla}_j)/(2i)$, 
$\mathbf{k}^\prime = -(\bm{\nabla}_i^\prime - \bm{\nabla}_j^\prime)/(2i)$,
and $P_\sigma$ is the spin-exchange operator.
Here, $W_0$ denotes the strength of the effective spin-orbit coupling, which is not active in homogeneous matter.
It, therefore, must be determined from nuclear data.
This is similar in form to a generalized Skyrme model proposed in Refs.~\cite{CBMBD03,CBBDM04,ADK06}, 
but the protocol for determining the Skyrme potential parameters is quite different.
In the so-called generalized Skyrme potential model, the parameters are determined by some properties of specific nuclei.
In our case, however, we will begin with an unchanged EoS and use very few nuclear data for remaining undetermined parameters. 
We also retain the freedom to have, e.g., $t_{33}=0$ but $y_{33}\neq 0$. 
The corresponding EDF in terms of the local densities as well as gradient and kinetic terms can be obtained from a standard calculation as
\begin{eqnarray}
\mathscr{E}
&=& \frac{\hbar^2}{2m}\tau
 +\frac{3}{8} t_0^{}\rho - \frac{1}{8} (t_0^{} + 2y_0^{}) {\rho}{\delta}^2  
\nonumber \\ & & \!\!\!
 +\frac{1}{16}\sum_{n=1}^{N-1}t_{3n}^{} \rho^{1+n/3} 
 \nonumber \\ & & \!\!\!
 - \frac{1}{48}\sum_{n=1}^{N-1}(t_{3n}^{} + 2y_{3n}^{})\rho^{1+n/3} \delta^2   
\nonumber \\ & & \!\!\!
 +\frac{1}{64}(9t_1^{} -5t_2^{} -4y_2^{}) \frac{(\nabla \rho)^2}{\rho}
\nonumber \\ & & \!\!\!
 -\frac{1}{64}(3t_1^{} + 6y_1^{} - t_2^{} - 2y_2^{} ) \frac{(\nabla\rho\delta)^2}{\rho}
\nonumber \\ & & \!\!\!
 +\frac{1}{8} ( 2t_1^{} + y_1^{}  + 2 t_2^{} + y_2^{} ) \tau
\nonumber \\ & & \!\!\!
 -\frac{1}{8}(t_1^{} + 2y_1^{}) -t_2^{} - 2y_2^{}) \sum_q \frac{\rho_q\tau_q}{\rho}
\nonumber \\ & & \!\!\!
 +\frac{1}{2} W_0 \left( \frac{{\mathbf J}\cdot\bm{\nabla}\rho}{\rho} 
 +\sum_q \frac{{\mathbf J}_q \cdot \bm{\nabla} \rho_q^{}}{\rho} \right),  
\label{eq:skyrmeedf}
\end{eqnarray}
where $\tau$ denotes the kinetic energy density and ${\mathbf J}$ the current density. 
The sum over $q$ means the summation over isospin, i.e., $q=(n,p)$.
Matching the KIDS EDF in Eq.~(\ref{eq:cdelta}) and the Skyrme functional in Eq.~(\ref{eq:skyrmeedf}) leads to the following relations:
\begin{eqnarray} \label{eq:relation} 
t_0^{} &=& \frac{8}{3} c_0^{} (0)  ,  \quad  
y_0^{}=\frac{8}{3} c_0^{}(0) - 4 c_0^{}(1),  
\nonumber \\  
t_{3n}^{} &=& 16 c_n^{} (0) \, , 
\quad y_{3n}^{} = 16 c_n^{} (0) - 24 c_n^{} (1), \quad (n \neq 2)  
\nonumber \\  
t_{32}^{} &=& 16 c_2^{} (0) - \frac{3}{5} \left(\frac{3}{2}\pi^2 \right)^{2/3} \theta_s 
\nonumber \\  
&\equiv& 16c_2(0)(1-\zeta ), 
\nonumber \\ 
y_{32}^{} &=& 16 c_2^{}(0) - 24c_2^{}(1) + \frac{3}{5}(3\pi^2)^{2/3} \left( 3\theta_{\mu} - \frac{\theta_s}{2^{2/3}}\right) 
\nonumber \\ 
&\equiv& [16c_2^{}(0)-24c_2^{}(1)](1-\zeta'),
\label{eq:tiyi} 
\end{eqnarray} 
which defines $\zeta$ and $\zeta'$ with
\begin{eqnarray} 
\theta_s &\equiv& 3t_1^{} + 5 t_2^{} + 4y_2^{} = \frac{5}{3}(3\pi^2)^{-2/3} 16c_2^{}(0)\zeta \nonumber \\ 
\theta_{\mu} &\equiv&  
t_1^{} + 3t_2^{} - y_1^{} + 3y_2^{} 
\nonumber \\ 
&=& \frac{\theta_s}{3\cdot 2^{2/3}}
\nonumber \\ && \mbox{}
-\frac{5}{9}(3\pi^2)^{-2/3} \left[16c_2^{}(0)-24c_2^{}(1) \right]\zeta’ .
\label{eq:theta} 
\end{eqnarray} 
The matching reveals that there are two sources for the $\rho^{5/3}$ term in the EoS which corresponds to $n=2$ in Eq.~(\ref{eq:skyrme}): 
one from the density-dependent terms in Eq.~(\ref{eq:skyrme}) with the Skyrme parameters $t_{32}^{}$, $y_{32}^{}$, and the other
from the momentum-dependent terms in Eq.~(\ref{eq:skyrme}) with the Skyrme parameters $t_1^{}$, $t_2^{}$, $y_1^{}$, $y_2^{}$. 
The partition is encoded in the unknown parameters $\zeta$ and $\zeta'$ in Eqs.~(\ref{eq:tiyi}) and (\ref{eq:theta}).  
Also undetermined at this point is the effective spin-orbit coupling strength $W_0$.

Following the simple procedure of Ref.~\cite{GPHPO16}, in the present work, we set $y_1^{} = y_2^{} = 0$ and 
assume $\zeta =\zeta'$, which leaves only 2 parameters, i.e., $\zeta$ and $W_0$, to be determined by nuclear data. 
In this case, the isoscalar and isovector effective mass parameters, $\mu_s\equiv m^*/m$ and $\mu_v\equiv m_v^*/m$, 
where $m$ denotes the nucleon mass in free space,
are not independent but are determined via $\zeta$ according to their relations to $\theta_s$ and $\theta_{\mu}$ as~\cite{CBHMS97} 
\begin{eqnarray} 
\mu_s^{-1}\equiv ({m^{\ast}_{\mathrm{}}/m})^{-1} &=& 1+\frac{m \rho}{8\hbar^2} \theta_s , 
\nonumber \\ 
\mu_v^{-1}\equiv ({m^{\ast}_v/m})^{-1} &=& 1+ \frac{m \rho}{4\hbar^2} (\theta_s-\theta_{\mu})  .
\label{eq:effm} 
\end{eqnarray}

A refined method taking full advantage of the momentum-dependent terms was developed and applied in Ref.~\cite{GPHO18}. 
The refinement was found inconsequential for bulk and static nuclear properties.
Therefore, the above simplified procedure with $y_1=y_2=0$ suffices for our present purpose. 
We now return to the issue of the expansion and examine whether three SNM terms and four PNM terms, a total of seven EoS parameters, 
are sufficient to achieve convergence of results in the case of nuclei as well as in homogeneous matter.

\section{Expansion in symmetric part
\label{symmetric}} 

Equipped with the formalism as discussed above, we now consider the issue of convergence in the description of nuclear properties. 
The question we address at this stage is how many terms are required for convergence of the expansion in Eq.~(\ref{eq1}); 
put in another way, at which order further EoS parameters, such as curvature or compressibility, and skewness, become inconsequential for nuclear applications 
and thus cannot be constrained by nuclear data. 
Specifically, we want to know whether higher accuracy can be achieved with more than three terms in SNM and more than four terms in PNM 
(or the symmetry energy) in practical applications. 
A negative answer would be of great importance since it would mean that the use of more terms can only lead to overfitting and risk loss of predictive power.  
The case of SNM will be investigated in this section and the next section is devoted to the case of PNM.

We proceed to examine whether variations in the value of $Q_0$ affect strongly the nuclear EoS and the quality of the description of nuclear structure. 
The empirically determined range of $Q_0$ value is between $-1200$~MeV and $-200$~MeV~\cite{FPT97}, which shows a huge uncertainty. 
An analysis of nuclear models provides a narrower range $-425.6 \sim -362.5$~MeV~\cite{DLSD12}, which still represents an uncertainty of the order of 15\%.
Taking this range as a reference, we choose three values of skewness coefficient, $-360$ MeV, $-390$~MeV, and $-420$~MeV. 
The four parameters $\alpha_{0,1,2,3}^{}$ are now determined by solving a $4\times 4$ system of equations where 
the coefficients are determined by the assumed values of $\rho_0^{}, E_0, K_0, Q_0$.

In the following, the sets of parameters resulting from $Q_0 =-360$, $-390$, and $-420$~MeV are labeled as S4a, S4b, and S4c, respectively,
with the number 4 referring to the number of expansion terms.
Presented in Table~\ref{table1} are the obtained values of the parameters $\alpha_i^{}$.
In this process, $c_i^{} (1)=\alpha_i^{} + \beta_i^{}$ in Eq.~(\ref{eq:cdelta}) are fixed to the KIDS-ad2 values of Ref.~\cite{PPLH16}, 
which parametrize the APR EoS for pure neutron matter. 
The $N=3$ case, model S3b is obtained with setting $\alpha_3^{} = 0$ but with $\alpha_3^{} + \beta_3^{} = 956.47$.%
\footnote{For consistency, the model with $N=3$ should be determined with $N=3$ parameterization for PNM. In fact, this model is equivalent to model P3
described in the next section. The final results for S3b and P3 are similar, but, in this section, we work with S3b to vary the SNM parameters only.} 
It can be found that the ranges of $\alpha_0^{}$ and $\alpha_1^{}$ are rather stable but those of higher order $\alpha_{2,3}^{}$ are sensitive to the input data.
Even the signs of the higher order parameters are not robust.
This uncertainty is expected because the input data are provided at nuclear saturation density and the higher order coefficients are influenced by higher- and lower-density
regions.
However, the resulting physical quantities of our interests are not so sensitive as will be shown below.

\begin{table}[tbp]
\renewcommand{\arraystretch}{1.1}
\begin{center}
\begin{tabular}{|c|c|cccc|} \hline\hline
Model &   $N$	& $\alpha_0^{}$	&	$\alpha_1^{}$	& $\alpha_2^{}$	&	$\alpha_3^{}$  \\ \hline
S3b &3		& $-664.52$ 		& $763.55$		& $40.13$ 		& $0$ 	\\
S4a 	&4	& $-677.69$ 		& $836.34$		& $-93.95$		& $82.33$  \\	
S4b  	&4	& $-646.44$ 		& $663.65$		& $224.15$		& $-112.99$ \\
S4c 	&4	& $-615.19$ 		& $490.96$		& $542.24$		& $-308.30$ 
 \\ 
		\hline\hline
 PNM &   $N$	& $\alpha_0^{} + \beta_0^{}$	&	$\alpha_1^{} + \beta_1^{}$	& 
 $\alpha_2^{} + \beta_2^{} $	&	$\alpha_3^{} + \beta_3^{} $  \\ \hline
KIDS-ad2 &4		& $-411.13$ 		& $1007.78$		& $-1354.64$ 		& $956.47$ 	\\
\hline \hline
\end{tabular}
\end{center}
\caption{Fitted values of parameters $\alpha_i^{}$ in units of $\si{MeV} \cdot \si{fm}^{3+i}$.
Model S3b with $N=3$ the EoS parameters are fixed assuming $\alpha_3 =0$ with $\rho_0^{} = 0.16$~fm$^{-3}$, $E_0 = -16.0$~MeV, and 
$K_0 = 240.0$~MeV with $\beta_i$ of KIDS-ad2.
Models S4a, S4b, and S4c correspond to $Q_0 = -360$, $-390$, and $-420$~MeV, respectively. 
For S3b, we obtain $Q_0 = -372.65$~MeV. 
The EoS of PNM is fixed by the baseline parameters shown at the bottom, which corresponds to KIDS-ad2.}
\label{table1}
\end{table}

\begin{figure*}[t]
\centering
\includegraphics[width=0.9\textwidth]{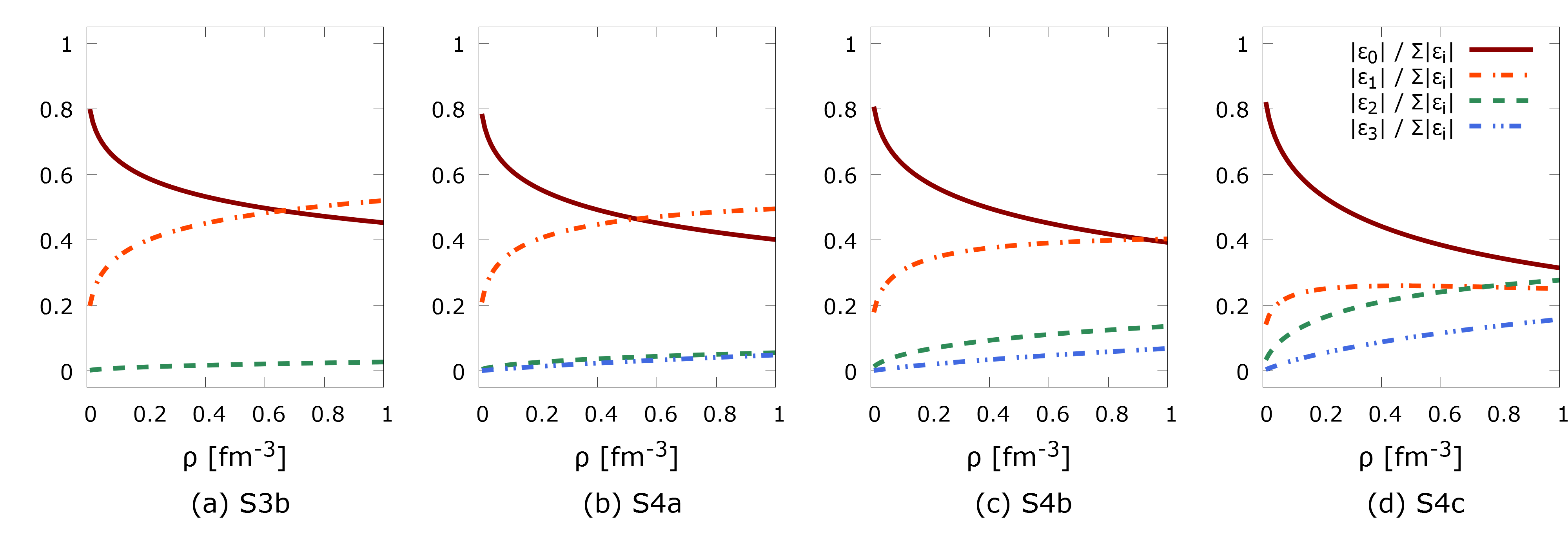}
\caption{
Relative magnitude of each interaction potential for symmetric matter for model
(a) S3b, (b) S4a, (c) S4b, and (d) S4c.}
\label{fig1} 
\end{figure*}

Figure~\ref{fig1} shows the relative magnitude of each interaction term $\varepsilon_i^{} = c_i^{} (0) \rho^{1+i/3} = \alpha_i^{} \rho^{1+i/3}$, 
namely, $|\varepsilon_i^{}| / \sum_i | \varepsilon_i^{}|$. 
The converging behavior $|\varepsilon_0^{}| > |  \varepsilon_1^{}| > | \varepsilon_2^{}| > | \varepsilon_3^{}|$
is satisfied well up to densities around $3 \rho_0^{}$ regardless of $N$ or $Q_0$ values.
At higher densities, where high-order terms are more active, the effects of varying $Q_0$ values become clearer as expected.
The dominance of the lowest-order term $\varepsilon_0^{}$ persists in all cases.

\begin{figure}[t]
\centering
\includegraphics[width=0.4\textwidth]{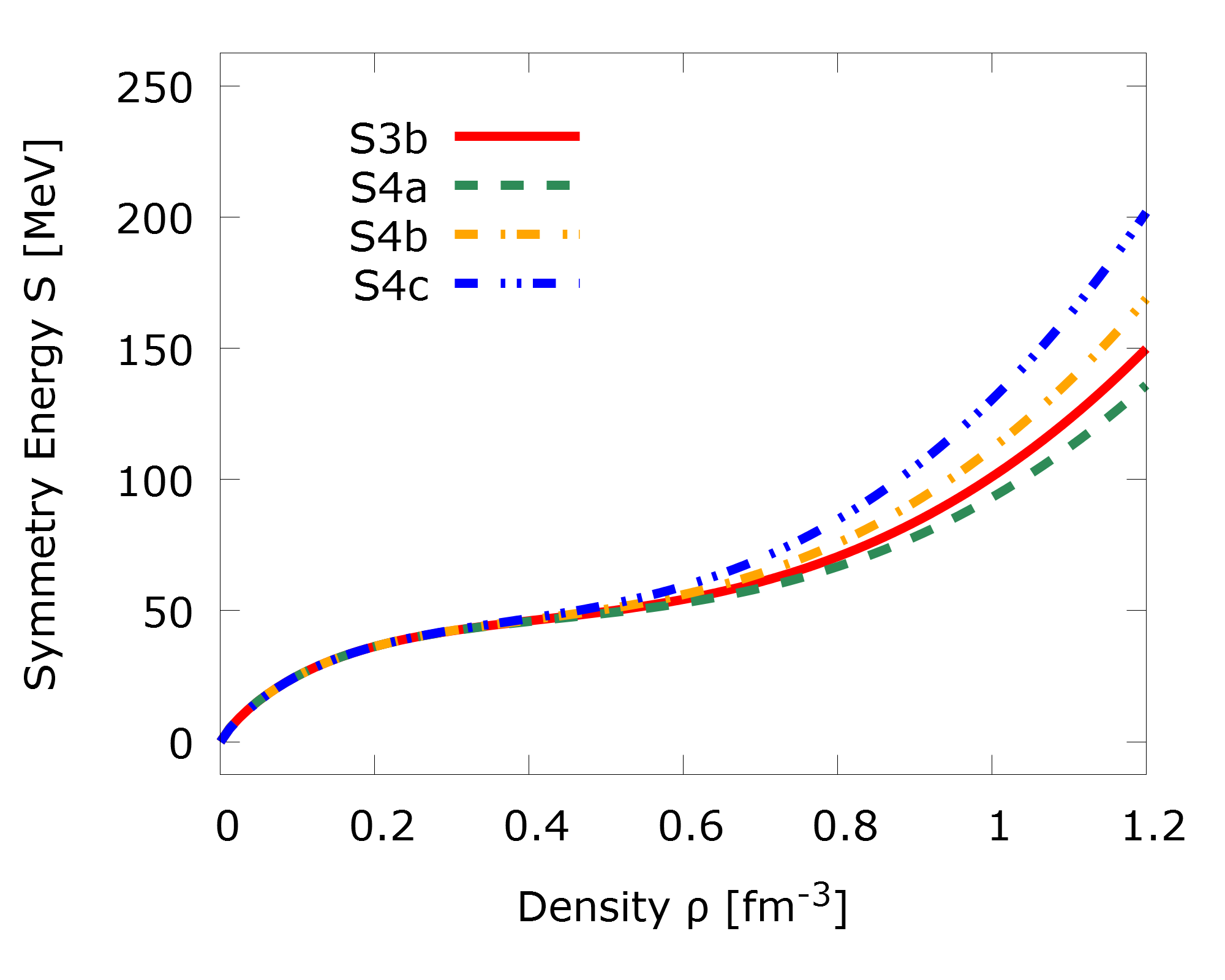}
\caption{Symmetry energy obtained with the parameter sets for symmetric nuclear matter from Table~\ref{table1}. 
The EoS of pure neutron matter is fixed to the baseline set KIDS-ad2.}
\label{fig2}
\end{figure}

Extrapolation of the model to higher densities is tested by considering properties of the neutron star. 
It is widely accepted that the core of a neutron star is very asymmetric, so the EoS of a neutron star could be sensitive to the symmetry 
energy in Eq.~(\ref{Eq:SymEn}) that is written in terms of $\beta_i$.
Since $c_i^{} (1) = \alpha_i^{} + \beta_i^{}$ is fixed by the parameter set KIDS-ad2, but $\alpha_i$ varies according to
$Q_0$ value, $\beta_i$ changes to keep $c_i(1)$ unchanged, and, consequently, $\mathcal{S}(\rho)$ depends on the $Q_0$ value.
Figure~\ref{fig2} shows the symmetry energy in various choices of $N$ and $Q_0$ values.
The dependence on $Q_0$ becomes more evident as density increases.
However, even around $0.8$~fm$^{-3}$ ($\sim 5 \rho_0^{}$), the maximum difference is only about 20~MeV.
The difference becomes appreciable as density reaches about 1~fm$^{-3}$, which is close to the maximum density
in the neutron star core and where, in any case, an EDF based on nucleonic degrees of freedom is questionable.

\begin{figure}[t]
\centering
\includegraphics[width=0.4\textwidth]{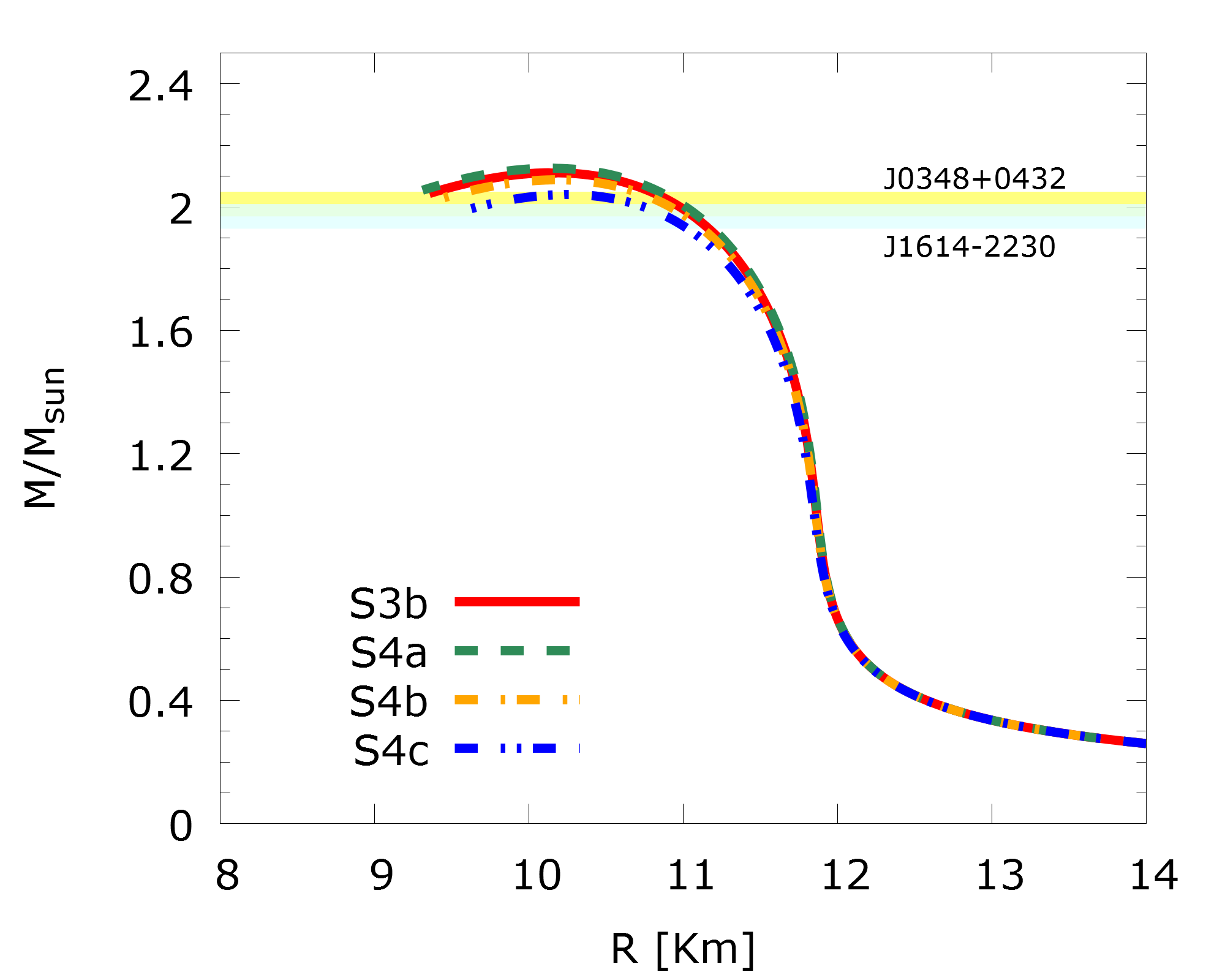}
\caption{Neutron star mass-radius relations: Results correspond to the respective symmetry-energy curves of Fig.~\ref{fig2}.
The bands are the range of neutron star masses reported in Refs.~\cite{DPRRH10,AFWT13}.}
\label{fig3}
\end{figure}

%
\begin{table*}[t]
\renewcommand{\arraystretch}{1.1}
\begin{center}
\begin{tabular}{c|ccccc|ccccc} \hline\hline
 \multirow{2}{*}{ Nucleus }
 			& \multicolumn{5}{c|}{Binding energy per nucleon $(E/A)$ (MeV)}						
			& \multicolumn{5}{c}{Charge radius $(R_c)$ (fm)}								\\
 			\cline{2-11}
 			& Expt. 	& S3b	& S4a	& S4b	& S4c
			 & Expt. 	& S3b	& S4a	& S4b	& S4c
 			\\ \hline
 \multirow{2}{*}{ \nuclide[40]{Ca} }
 			& $8.5513^*$	& $8.5565$ 		& $8.5579$		& $8.5544$		& $8.5512$
			& $3.4776^*$	& $3.4781$ 		& $3.4799$		& $3.4758$		& $3.4720$ 
 			\\
 			&			& $(0.060\%)$		& $(0.078\%)$		& $(0.037\%)$		& $(0.001\%)$
			&			& $(0.014\%)$		& $(0.066\%)$		& $(0.052\%)$		& $(0.161\%)$
			\\ 
 \multirow{2}{*}{ \nuclide[48]{Ca} }
			& $8.6667^*$	& $8.6564$ 		& $8.6569$		& $8.6558$		& $8.6549$ 
			& $3.4771^*$	& $3.4867$		& $3.4882$		& $3.4847$		& $3.4813$
 			\\
 			&			& $(0.120\%)$		& $(0.113\%)$		& $(0.126\%)$		& $(0.136\%)$
			&			& $(0.277\%)$		& $(0.319\%)$		& $(0.220\%)$		& $(0.122\%)$
			\\ 
 \multirow{2}{*}{ \nuclide[208]{Pb} }
			& $7.8675^*$	& $7.8809$		& $7.8816$		& $7.8800$		& $7.8783$ 
			& $5.5012^*$	& $5.4887$		& $5.4901$		& $5.4870$		& $5.4840$
 			\\
 			&			& $(0.172\%)$		& $(0.179\%)$		& $(0.160\%)$		& $(0.138\%)$
			&			& $(0.228\%)$		& $(0.201\%)$		& $(0.259\%)$		& $(0.313\%)$
 			\\
 			\hline\hline
 \multirow{2}{*}{ \nuclide[16]{O} }
 			& $7.9762$	& $7.8684$ 		& $7.8675$ 		& $7.8686$		& $7.8678$	
			& $2.6991$	& $2.7618$ 		& $2.7643$		& $2.7587$		& $2.7541$
 			\\
 			&			& $(1.35\%)$		& $(1.36\%)$		& $(1.35\%)$		& $(1.36\%)$
			&			& $(2.322\%)$		& $(2.41\%)$		& $(2.209\%)$		& $(2.036\%)$
			\\ 
 $^{28}$O		& ---			& $6.0646$		& $6.0640$		& $6.0650$		& $6.0649$
 			& ---			& $2.8371$ 		& $2.8384$		& $2.8351$		& $2.8315$
 			\\ 
 $^{60}$Ca	& --- 			& $7.6561$		& $7.6567$		& $7.6552$		& $7.6535$ 
 			& ---			& $3.6465$		& $3.6478$		& $3.6445$		& $3.6411$
			\\ 
 \multirow{2}{*}{ \nuclide[90]{Zr} }
			& $8.7100$	& $8.7328$		& $8.7345$		& $8.7309$		& $8.7282$ 
			& $4.2694$	& $4.2476$		& $4.2488$		& $4.2459$		& $4.2428$
 			\\
 			&			& $(0.263\%)$		& $(0.281\%)$		& $(0.241\%)$		& $(0.209\%)$
			&			& $(0.510\%)$		& $(0.482\%)$		& $(0.550\%)$		& $(0.622\%)$
 			\\ 
 \multirow{2}{*}{ \nuclide[132]{Sn} }
  			& $8.3549$	& $8.3563$ 		& $8.3559$		& $8.3565$		& $8.3565$ 
			& $4.7093$	& $4.7089$ 		& $4.7100$		& $4.7072$		& $4.7044$
 			\\
 			&			& $(0.017\%)$		& $(0.013\%)$		& $(0.020\%)$		& $(0.020\%)$
			&			& $(0.009\%)$		& $(0.015\%)$		& $(0.044\%)$		& $(0.103\%)$
			\\ \hline\hline
\end{tabular}
\end{center}
\caption{Binding energies per nucleon and charge radii of selected spherical magic nuclei computed with four EoS parameter sets. 
Top three values with an asterisk for \nuclide[40]{Ca}, \nuclide[48]{Ca}, and \nuclide[208]{Pb} represent input data and the others are predictions. 
Numbers in parentheses denote the percentage deviations of predictions from data.
Experimental data are from Refs.~~\cite{NNDC,AM13}.}
\label{tab:nucl1}
\end{table*}

%
\begin{table*}[t]
\renewcommand{\arraystretch}{1.1}
\begin{center}
\begin{tabular}{c|cccc} \hline\hline
Parameter &  S3b   & S4a & S4b & S4c \\ \hline
$t_0^{}$ ($\mbox{MeV} \cdot \mbox{fm}^3$) & $-1772.04$ & $-1807.17$ & $-1723.84$ & $-1640.50$ \\
$y_0^{}$ ($\mbox{MeV} \cdot \mbox{fm}^3$) &  $-127.52$ & $-162.65$ & $-79.32$ & $4.02$ \\
$t_1^{}$ ($\mbox{MeV} \cdot \mbox{fm}^5$) & $275.72$ & $262.17$ & $288.94$ & $303.28$ \\
$t_2^{}$ ($\mbox{MeV} \cdot \mbox{fm}^5$) & $-161.50$ & $-167.94$ & $-154.90$ & $-146.98$ \\
$t_{31}^{}$ ($10^{4} \, \mbox{MeV} \cdot \mbox{fm}^4$) & $1.222 $ & $1.338 $ & $1.062 $ & $0.7855 $ \\
$y_{31}^{}$ ($10^{4} \, \mbox{MeV} \cdot \mbox{fm}^4$) & $-1.197$ & $-1.081$ & $-1.357$ & $-1.633$ \\
$t_{32}^{}$ ($ \mbox{MeV} \cdot \mbox{fm}^5$) & $571.0$ & $-1310.7$ & $3252.4$ & $8043.0$  \\
$y_{32}^{}$  ($ 10^{4} \,  \mbox{MeV} \cdot \mbox{fm}^5$) & $2.949$ & $2.704$ & $3.274$ & $3.818$ \\
$t_{33}^{}$ ($\mbox{MeV} \cdot \mbox{fm}^6$) & --- & $1317.2$ & $-1807.8$ & $-4932.8$  \\
$y_{33}^{}$ ($10^4 \, \mbox{MeV} \cdot \mbox{fm}^6$)  & $-2.296$ & $-2.164$ & $-2.476$ & $-2.789$ 
\\ \hline
$\zeta$ & 0.1106 & 0.1281 & 0.0931 & 0.0729 \\
$W_0$ ($\mbox{MeV}\cdot \mbox{fm}^5$) & 108.35 & 106.79 & 109.88 & 111.55 
\\
 \hline \hline
\end{tabular}
\end{center}
\caption{Fitted parameters of Skyrme functional parameters. 
Here, we set $y_1^{} = y_2^{} = 0$ and $\zeta$ is dimensionless.}
\label{tab:SNM}
\end{table*}

The difference in model predictions on the mass-radius relation of neutron stars is shown in Fig.~\ref{fig3}.
Predictions for the maximum neutron star mass are 2.11, 2.13, 2.09, and 2.04~$M_\odot$, where $M_\odot$ is solar mass, 
for S3b, S4a, S4b, and S4c, respectively. 
This shows that all the four parameter sets give similar mass-radius properties of neutron stars and allow $2 M_\odot$ as a neutron star mass.
This observation indicates that the effect of fourth order $\varepsilon_3^{}$ term and, in particular, the variation of the $Q_0$ value within the range of
Ref.~\cite{DLSD12} is marginal in the considered physical quantities.

\begin{table*}[t]
\renewcommand{\arraystretch}{1.1}
\begin{center}
\begin{tabular}{c|c|ccccccc|ccccc} \hline\hline
Model & $N$	& $c_0^{}(1)$	& $c_1^{}(1)$	& $c_2^{}(1)$	& $c_3^{}(1)$ & $c_4^{}(1)$	& $c_5^{}(1)$ & $\chi_n^2$ 
& $J$ & $L$ & $K_{\rm sym}$ & $Q_{\rm sym}$ & $R_{\rm sym}$ \\ \hline	
P3 &	3	& $-266.72$		& $133.50$		& $281.38$		& ---	
 		& ---				& --- 				& $5.3 \times 10^{-4}$		
		& $32.6$ & $53.5$ & $-129.7$ & $422.3$ & $-2421.8$ 		
 		\\
P4 &  4	& $-407.94$ 		& $990.09$		& $-1321.86$		& $937.14$	
 		& ---				& ---				& $1.4 \times 10^{-4} $	
		& $32.8$ & $49.2$ & $-156.3$ & $583.1$ & $-2469.7$
 		\\	
P5 &  5	& $-224.16$ 		& $-479.28$		& $2814.48$		& $-3963.71$	
 		& $2075.79$		& ---				& $6.3 \times 10^{-5} $		
		& $33.0$ & $51.4$ & $-166.8$ & $461.4$ & $-1388.4$ 
		\\ \hline
P6a &  6	& $-224.81$ 		& $-473.46$		& $2795.50$		& $-3935.18$	
 		& $2056.11$		& $4.94$		& $6.3 \times 10^{-5}$		
		& $33.0$ & $51.4$ & $-166.8$ & $461.6$ & $-1391.7$ \\
P6b &  6	& $-283.99$ 		& $110.63$		& $604.05$		& $-10.59$	
 		& $-1312.44$		& $1117.76$		& $6.4 \times 10^{-5}$		
		& $33.0$ & $51.5$ & $-163.8$ & $450.0$ & $-1545.9$
 		\\
P6c &  6	& $-313.98$ 		& $400.88$		& $-463.41$		& $1864.00$	
 		& $-2891.61$		& $1630.37$		& $6.5 \times 10^{-5}$	
		& $33.0$ & $51.5$ & $-162.3$ & $446.6$ & $-1631.2$	
               \\ \hline \hline 
\end{tabular}
\end{center}
\caption{ 
Values of $c_i(^{}1)$ fitted to APR EoS of PNM. 
The unit of $c_i$ is $\si{MeV} \cdot \si{fm}^{3+i}$ and the units of $J$, $L$, $K_{\rm sym}$, $Q_{\rm sym}$, and $R_{\rm sym}$ are MeV.}
\label{tab:PNM}
\end{table*}

\begin{table*}[t]
\renewcommand{\arraystretch}{1.1}
\begin{center}
\begin{tabular}{c|c|ccccccc|ccccc} \hline\hline
Model & $N$	& $c_0^{}(1)$	& $c_1^{}(1)$	& $c_2^{}(1)$	& $c_3^{}(1)$ & $c_4^{}(1)$	& $c_5^{}(1)$ & $\chi_n^2$ 
& $J$ & $L$ & $K_{\rm sym}$ & $Q_{\rm sym}$ & $R_{\rm sym}$ \\ \hline
QMC P3 & 3 & $-119.01$ & $-424.80$ & $841.72$ & --- 
   		&  --- & --- & $1.4 \times 10^{-4}$ 
		& $34.1$ & $62.5$ & $-59.5$ & $566.0$ & $-3304.7$
		\\
QMC P4  & 4  &  $-395.79$           & $1085.08$           & $-1818.25$         &  $ 1519.80$ 
               & ---                       & ---                      & $1.5\times 10^{-6}$ 
               & $34.5$ & $60.5$ & $-88.9$ & $751.2$ & $-3075.9$ 
              \\   
QMC P5  & 5  &   $-349.56$         & $740.06$              & $-876.85 $          &  $404.60$ 
              &   $484.89$           & ---                    & $1.3\times 10^{-6}$ 
              & $34.5$ & $60.7$ & $-90.0$ & $735.7$ & $-2876.4$
               \\ 
QMC P6  & 6  &   $-293.93$         & $238.03$              & $896.89 $          &  $-2667.81$ 
              &   $3098.48$           & $-874.88$                    & $1.3\times 10^{-6}$ 
              & $34.5$ & $60.7$ & $-90.8$ & $737.1$ & $-2781.0$
               \\ \hline \hline 
\end{tabular}
\end{center}
\caption{ 
Same as Table~\ref{tab:PNM} but for QMC EoS of PNM of Ref.~\cite{CGPP14}. }
\label{tab:PNM-QMC}
\end{table*}

Now we extend our investigation to the structure of finite nuclei.
In order to make use of the Kohn-Sham framework, it is most convenient to
transform the EDF to the form of a Skyrme potential, as described in Sec.~\ref{Sec:Skyrme}.
When we expand the EDF up to $N=4$, we have $2N$ parameters that are determined from the bulk properties of homogeneous SNM and PNM. 
However, the effective Skyrme interaction of Eq.~(\ref{eq:skyrme}) has five additional parameters.
With the assumption that $y_1^{} = y_2^{} = 0$ and $\zeta = \zeta'$, two parameters, $\zeta$ and $W_0$, are yet to be determined.
The fitting of the undetermined parameters $\zeta$ and $W_0$ is performed using 6 data points, namely, the energy per particle ($E/A$) 
and charge radius ($R_c$) of \nuclide[40]{Ca} \nuclide[48]{Ca}, and \nuclide[208]{Pb}. 
These are listed in the upper 3 rows in Tables~\ref{tab:nucl1} and the fitted values of the Skyrme functional parameters are listed in 
Table~\ref{tab:SNM} for models S3b, S4a, S4b, and S4c.
We find that the uncertainties in $c_{i}^{}$ are mostly transferred into those in $t_{32}^{}$ and $t_{33}^{}$, and even their signs
change depending on the model.
However, the derived physical quantities of the considered nuclei are rather robust.
The resulting effective masses $\mu_s$ and $\mu_v$ of Eq.~(\ref{eq:effm}) are obtained as $\mu_s = 0.99$, 1.03, 0.96, and 0.92, while
$\mu_v = 0.82$, 0.85, 0.79, and 0.77 for S3b, S4a, S4b, and S4c, respectively.%
We emphasize again that the effective mass values can be allowed to vary, if desired, with no deterioration of the quality 
of the results on the considered nuclear data. More detailed discussion can be found in Ref.~\cite{GPHO18}.

The results for \nuclide[16]{O}, \nuclide[28]{O}, \nuclide[60]{Ca}, \nuclide[90]{Zr}, and \nuclide[132]{Sn} are also given in Tables~\ref{tab:nucl1} 
for each model. 
For both $E/A$ and $R_c$, fitting quality and predictions of S4a, S4b and S4c are similar, and it is hard to distinguish these models.
Furthermore, it is also found that their predictions are similar to those of S3b, which means that the model with $N=3$ is sufficient enough 
in practical calculations.
This result leads to the conclusion that the three leading terms in the isospin symmetric part of the EDF are sufficient to describe not only the bulk properties 
of neutron stars but also magic nuclei. 
Both types of systems exhibit the same convergence behavior in a single and unified framework.

\section{Expansion in asymmetric part 
\label{asymmetric}}

In this section, we focus on the EDF expansion in asymmetric nuclear matter.
We perform this examination by retaining the KIDS-ad2 parameterization ($N=3$) for SNM, which was shown to be sufficient 
in the description of symmetric matter. 
With this constraint we proceed to examine the expansion behavior in PNM by varying EoS parameters. 
In Ref.~\cite{PPLH16} it was found that at least four terms are needed for satisfactory description of PNM or nuclear symmetry energy. 
In the present work, we increase the order of expansion of isospin asymmetric part from $N=3$ to $N=6$
and use the APR PNM EoS as input data because of lack of data for PNM.
Note that  the APR pseudodata are not smooth but show a kink at roughly twice the saturation density.  
Therefore, as in Ref.~\cite{PPLH16}, we assign a weight to the data at low energies by defining cost function 
$\chi^2$ as
\begin{eqnarray}
\chi^2 = \sum_j \exp \left(- \tilde\beta \rho_j^{} / \rho_0^{} \right)  \left( \frac{\mathscr{E}(\rho_j^{}) - D_j}{\mathscr{T}(\rho_j^{})} \right)^2,
\end{eqnarray}
where $D_j$ is the data point for density $\rho_j^{}$, $\mathscr{E}$ and $\mathscr{T}$ are the nuclear EDF and its the kinetic term given by Eq.~(\ref{eq1}), 
respectively, and we set $\tilde\beta =1$. 
We refer the details on this form to Ref.~\cite{PPLH16}.

\begin{figure*}[t]
\centering
\includegraphics[width=0.68\textwidth]{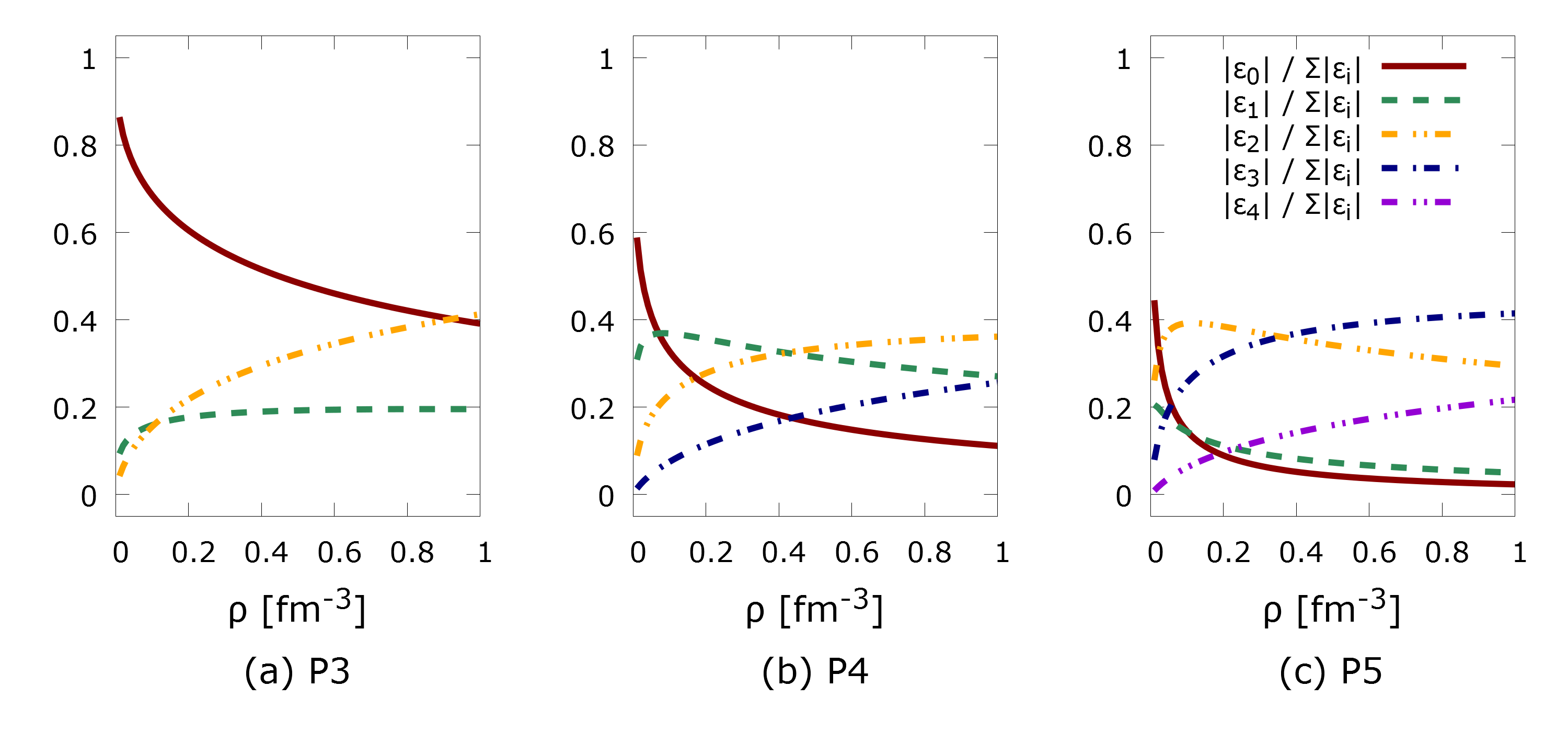}
\caption{Relative magnitude of each interaction potential for symmetric matter for model (a) P3, (b) P4, and (c) P5.} 
\label{fig4}
\end{figure*}

Fitted values of parameters and the corresponding $\chi_n^2$ defined as~\cite{PPLH16}
\begin{equation}
\chi_n^2 = \chi^2 / \sum_j {\rm exp} \left(- \tilde\beta \rho_j^{} / \rho_0^{} \right) 
\end{equation}
are shown in Table~\ref{tab:PNM}.
They are referred to as model P$N$ for $N=(3,4,5,6)$.
For $N=6$, we find that there may be more than two sets of parameters that have similar low $\chi^2$ values.
As examples, we give three sets, P6a, P6b, P6c in Table~\ref{tab:PNM}.
In particular, P6a is practically equal to P5 and it does not have any physical meaning to work with $N=6$ or higher for APR pseudodata.
This is expected since the APR EoS for PNM is determined at densities which can hardly be probed by higher-order terms.
The EoS parameters computed for each model are also shown in Table~\ref{tab:PNM}.
It can be found that although values of model parameters $c_i^{}$ would heavily depend on model, the resulting physical quantities or EoS parameters,
$J$, $L$, $K_{\rm sym}$, and even $Q_{\rm sym}$ are similar except $R_{\rm sym}$ that depends on the high-order behavior of EDF.
We also carry out this kind of analyses with the quantum Monte Carlo (QMC) results of Ref.~\cite{CGPP14} that are obtained with the AV8'+UIX interaction 
and verify this observation. 
The results are presented in Table~\ref{tab:PNM-QMC}.
In this case, we find that there may be more than two sets with similar accuracy even with $N=5$, although we do not list them in Table~\ref{tab:PNM-QMC}.
The fit quality hardly improves in  P6.
We therefore continue our investigation with models P3, P4, and P5 for the PNM parameters for further exploration.

We first plot the relative magnitudes of individual interaction terms for PNM in Fig.~\ref{fig4} for models P3, P4, and P5.
A common aspect in all three cases is the suppression of the $\varepsilon^{}_0$ term at high densities.
In particular, in P4 and P5, this suppression starts to happen already at the nuclear saturation density as pointed out in 
Ref.~\cite{PPLH16}. 
This behavior is different from that of the SNM case and this would indicates sophisticated dynamics in PNM, which
would imply nontrivial isospin dependence of dynamics in nuclear matter and causes huge theoretic uncertainties in nuclear symmetry energy.

\begin{figure}[t]
\centering
\includegraphics[width=0.4\textwidth]{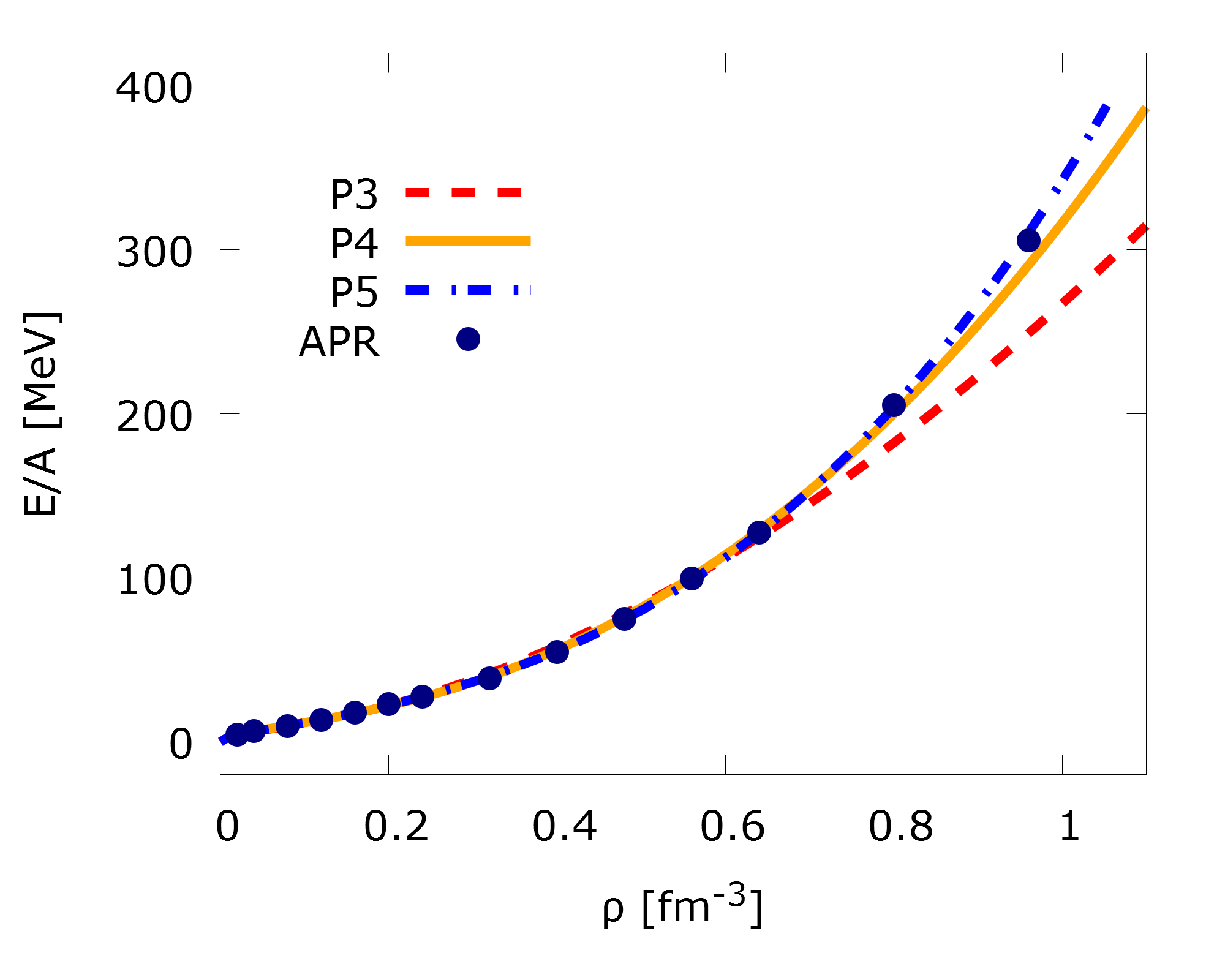}
\caption{Energy per particle of pure neutron matter with models P3, P4 and P5 presented in Table~\ref{tab:PNM}.
Here, the symmetric EoS parameters $\alpha_i^{}$ are fixed as model S3b in Table~\ref{table1}.}
\label{fig5}
\end{figure}

Figure~\ref{fig5} shows the energy per particle of PNM for each model and the obtained results are compared to the APR pseudodata.
This evidently shows that in order to reproduce the APR pesudodata up to high density region, we need at least $N=4$.
And it also shows that $N=5$ does not give any noticeable change from the result of P4.

\begin{figure}[t]
\centering
\includegraphics[width=0.4\textwidth]{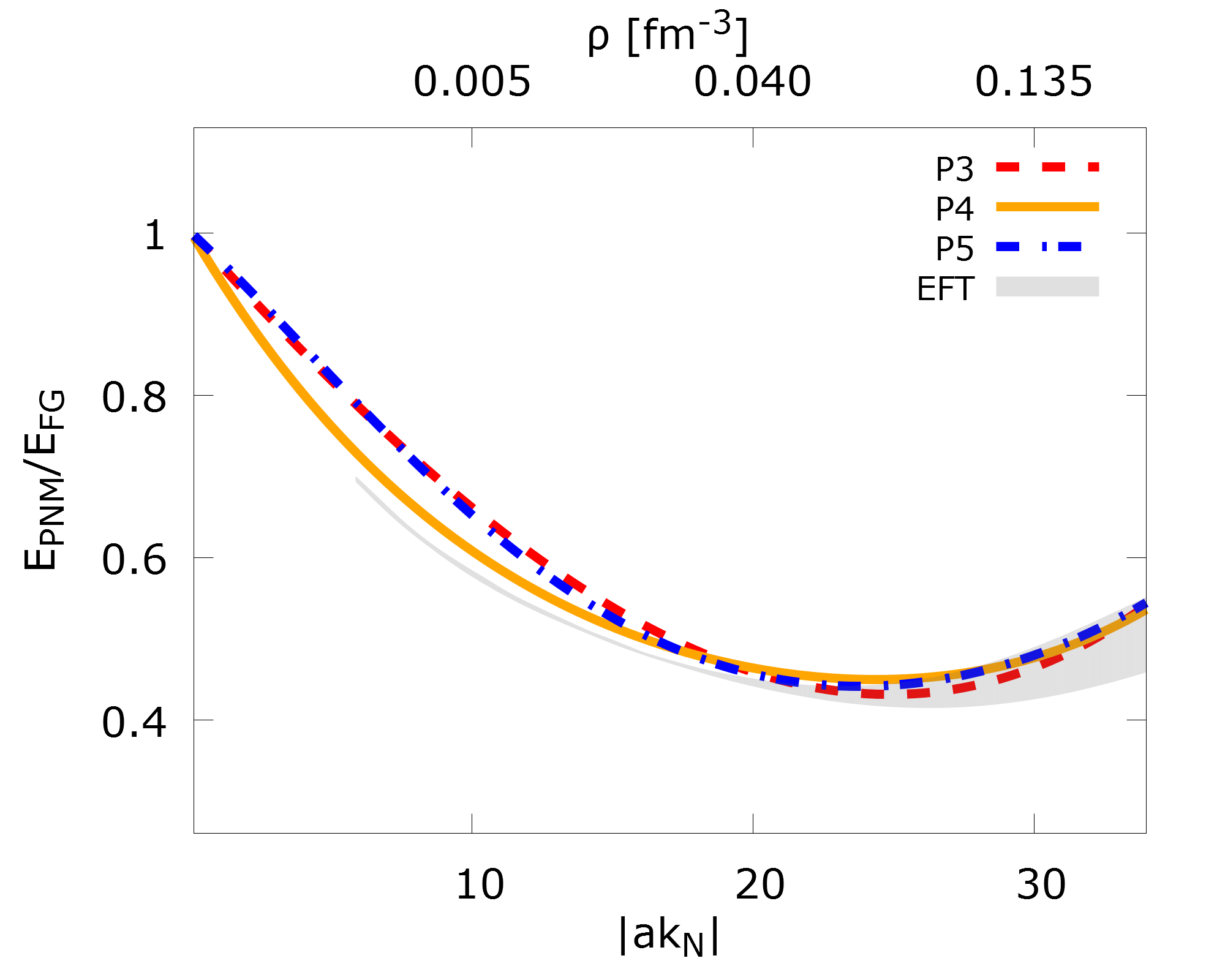}
\caption{Energy of pure neutron matter $E_{\rm PNM}$ divided by the free gas energy $E_{\rm FG}$
is compared to EFT results of Ref.~\cite{DSS13} at low densities, where $a(= -18.9~\mbox{fm})$ is the 
neutron-neutron scattering length in free space and $k_N^{}$ is the neutron Fermi momentum.
}
\label{fig6}
\end{figure}

\begin{figure}[t]
\centering
\includegraphics[width=0.4\textwidth]{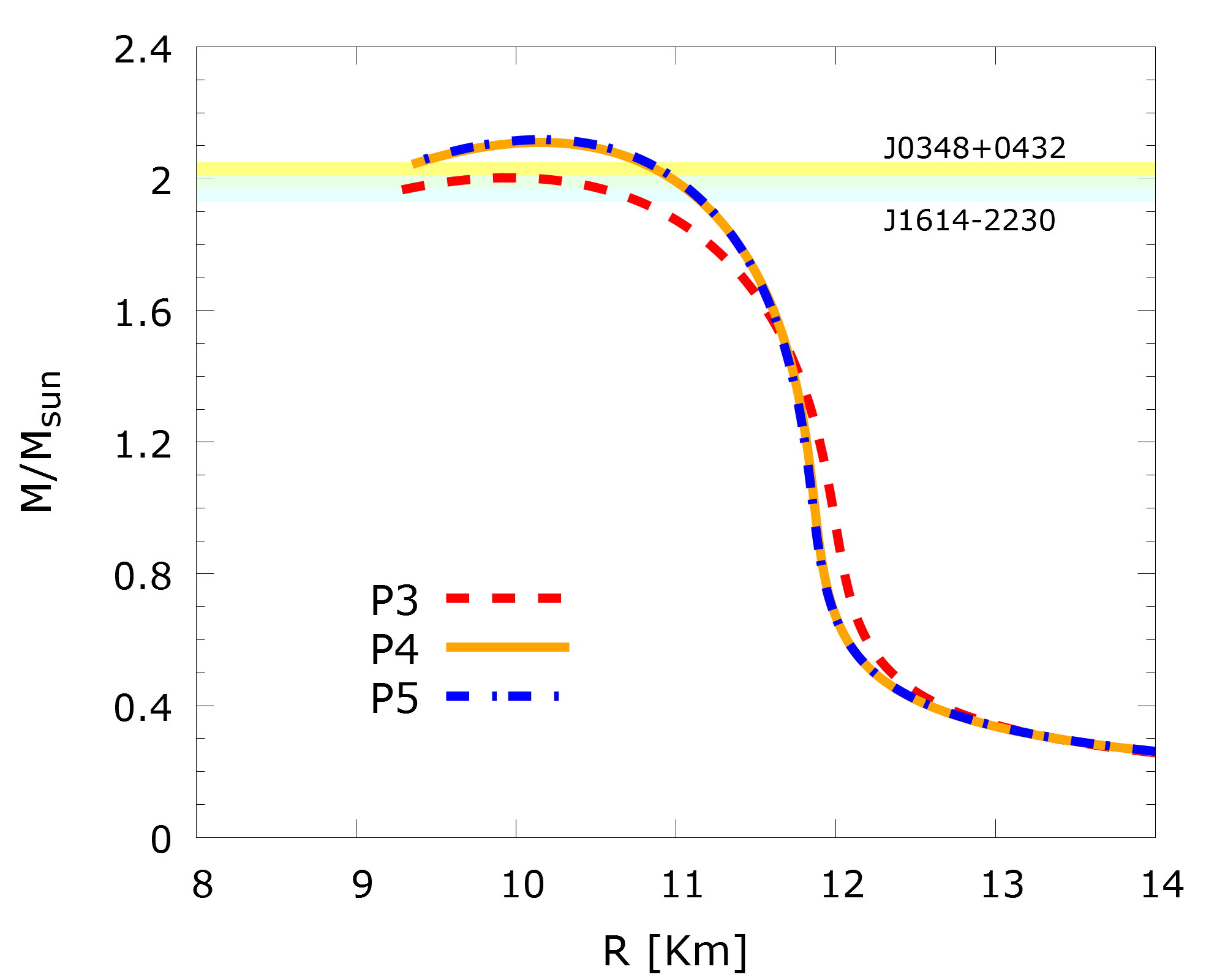}
\caption{Neutron star mass-radius relations for the parameter sets P3, P4, and P5.}
\label{fig7}
\end{figure}

In Fig.~\ref{fig6}, we present the energy per particle of PNM ($E_{\rm PNM}$) divided by the free gas energy ($E_{\rm FG}$)
at very low densities.
Effective field theory (EFT) results of Ref.~\cite{DSS13} are presented for comparison by a shaded band. 
Again, the good agreement with the EFT results is achieved with P4 and higher order terms are irrelevant. 
The irrelevance of higher order terms of EDF at low densities is not surprising but it is worthwhile to note that the parameters fitted at saturation point
can reproduce the results for very dilute system, which is a nontrivial result.
Because P4 and P5 have similar EoS, the corresponding neutron star mass-radius curves are expected to be similar and 
this is confirmed by the results shown in Fig.~\ref{fig7}.
Here again, the maximum neutron star mass is around $2\, M_\odot$.

%
\begin{table*}[t]
\renewcommand{\arraystretch}{1.1}
\begin{center}
\begin{tabular}{c|ccc} \hline\hline
Parameter & P3   & P4 & P5  \\ \hline
$t_0^{}$ ($\mbox{MeV} \cdot \mbox{fm}^3$) &  $-1772.04$	&	$-1772.04$	&	$-1772.04$ \\
$y_0^{}$ ($\mbox{MeV} \cdot \mbox{fm}^3$) & $-705.16$		&  $-140.27$	&	$-875.42$  \\
$t_1^{}$ ($\mbox{MeV} \cdot \mbox{fm}^5$) &  $247.33$		&  $275.83$	&	$269.90$ \\
$t_2^{}$ ($\mbox{MeV} \cdot \mbox{fm}^5$) & $-173.00$		& $-161.48$	&	$-163.95$\\
$t_{31}^{}$ ($10^{4} \, \mbox{MeV} \cdot \mbox{fm}^4$) &  $12216.73$	&	$12216.73$	&	$12216.73$\\
$y_{31}^{}$ ($10^{4} \, \mbox{MeV} \cdot \mbox{fm}^4$) &  $9012.81$	&	$-11545.41$	&	$23719.36$\\
$t_{32}^{}$ ($ \mbox{MeV} \cdot \mbox{fm}^5$) & $1087.14$	&	$569.38$ &	$678.46$   \\
$y_{32}^{}$  ($ 10^{4} \,  \mbox{MeV} \cdot \mbox{fm}^5$) & $-10346.18$	&	$28700.54$	&	$-70692.70$ \\
$y_{33}^{}$ ($10^4 \, \mbox{MeV} \cdot \mbox{fm}^6$)  & --- & $-22491.36$	&	$95128.93$ \\
$y_{34}^{}$ ($10^4 \, \mbox{MeV} \cdot \mbox{fm}^7$)  & --- & --- & $-49818.87$
\\ \hline
$\zeta$ &  $-0.6931$	 &	$0.1133$	&	$-0.0566$ \\
$W_0$ ($\mbox{MeV}\cdot \mbox{fm}^5$) &  $104.12$	&	$108.46$	&	$108.25$
\\
 \hline \hline
\end{tabular}
\end{center}
\caption{Same as Table~\ref{tab:SNM} but for P3, P4, and P5. Note that $t_{33}^{} = t_{34} = 0$ as we use S3b for $\alpha_i^{} = c_i^{}(0)$.}
\label{tab:PNM-1}
\end{table*}

\begin{table*}[t]
\renewcommand{\arraystretch}{1.1}
\begin{center}
\begin{tabular}{c|cccc|cccc} \hline\hline
 \multirow{2}{*}{ Nuclei{\;} }
 			& \multicolumn{4}{c|}{Energy per particle (MeV)}		
 			& \multicolumn{4}{c}{Charge radius (fm)}	\\
 			\cline{2-9}
 			& Expt. 		& P3			& P4			& P5			
			 & Expt. 		& P3			& P4			& P5		
 			\\ \hline
 \multirow{2}{*}{ $^{40}$Ca }
 			& $8.5513^*$	& $8.5573$ 		& $8.5564$		& $8.5561$	
			& $3.4776^*$	& $3.4785$ 		& $3.4781$		& $3.4782$	
 			\\
 			&			& $(0.070\%)$		& $(0.059\%)$		& $(0.056\%)$	
			&			& $(0.026\%)$		& $(0.014\%)$		& $(0.015\%)$	
			\\ 
 \multirow{2}{*}{ $^{48}$Ca }
			& $8.6667^*$	& $8.6556$ 		& $8.6565$		& $8.6581$		
			& $3.4771^*$	& $3.4891$		& $3.4867$		& $3.4870$
 			\\
 			&			& $(0.129\%)$		& $(0.118\%)$		& $(0.099\%)$		
			&			& $(0.345\%)$		& $(0.277\%)$		& $(0.285\%)$
                        \\ 
 \multirow{2}{*}{ $^{208}$Pb }
			& $7.8675^*$	& $7.8849$		& $7.8806$		& $7.8793$	
			& $5.5012^*$	& $5.4934$		& $5.4886$		& $5.4891$	
 			\\
 			&			& $(0.222\%)$		& $(0.167\%)$		& $(0.151\%)$		
			&			& $(0.141\%)$		& $(0.228\%)$		& $(0.221\%)$
 			\\
 			\hline\hline
 \multirow{2}{*}{ $^{16}$O }
 			& $7.9762$	& $7.8641$ 		& $7.8683$ 		& $7.8669$
			& $2.6991$	& $2.7634$ 		& $2.7618$		& $2.7621$		
 			\\
 			&			& $(1.405\%)$		& $(1.353\%)$		& $(1.371\%)$	
			&			& $(2.382\%)$		& $(2.322\%)$		& $(2.335\%)$	
			\\ 
 $^{28}$O		& ---			& $6.0705$		& $6.0628$		& $6.0585$	
 	 		& ---			& $2.8435$ 		& $2.8371$		& $2.8396$
 			\\ 
 $^{60}$Ca	& ---			& $7.6659$		& $7.6548$		& $7.6513$	
 			& ---			& $3.6511$		& $3.6465$		& $3.6478$
			\\ 
 \multirow{2}{*}{ $^{90}$Zr }
			& $8.7100$	& $8.7336$		& $8.7330$		& $8.7344$	
			& $4.2694$	& $4.2489$		& $4.2476$		& $4.2476$	
 			\\
 			&			& $(0.272\%)$		& $(0.264\%)$		& $(0.280\%)$	
			&			& $(0.480\%)$		& $(0.510\%)$		& $(0.511\%)$	
 			\\ 
 \multirow{2}{*}{ $^{132}$Sn }
  			& $8.3549$	& $8.3592$ 		& $8.3559$		& $8.3549$	
			& $4.7093$	& $4.7133$ 		& $4.7088$		& $4.7090$	
 			\\
 			&			& $(0.052\%)$		& $(0.013\%)$		& $(0.001\%)$	
			&			& $(0.085\%)$		& $(0.010\%)$		& $(0.006\%)$	
			\\ \hline\hline
\end{tabular}
\end{center}
\caption{Same as Table~\ref{tab:nucl1} but for P3, P4, and P5.
The SNM parameters are fixed to the values of model S3b in Table~\ref{table1}. The experimental data are from Refs.~ \cite{NNDC,AM13}.}
\label{tab:nuclei2}
\end{table*}

From the investigation for infinite nuclear matter properties and neutron star mass-radius relations, we conclude that at least four terms are necessary 
for reasonable descriptions.
Then the next question would be whether the parameters determined in this way can describe nuclear properties.
Here, we follow the same method and procedure used in Sect.~\ref{symmetric}.
The obtained Skyrme parameters are displayed in Table~\ref{tab:PNM-1}, which lead to the binding energy per nucleon and charge radius
as presented in Table~\ref{tab:nuclei2}.
This shows that there is no significant difference among the predictions of the three parameter sets and even P3 can give a reasonable description of
nuclear properties considered in the present work.%
We also performed these calculations with the parameter sets QMC P3, QMC P4, and QMC P5 listed in Table~\ref{tab:PNM-QMC} and they
lead to very similar results and conclusions.

\begin{table*}[t]
\renewcommand{\arraystretch}{1.1}
\begin{center}
\begin{tabular}{c|c|ccccc} \hline\hline
Model 	&  $R_{\rm sym}$	& $c_0^{}(1)$	&	$c_1^{}(1)$		& $c_2^{}(1)$		&	$c_3^{}(1)$ 	& $c_4^{}(1)$ \\ \hline
P5a 		& $-2170$				& $-329.19$		& $411.12$		& $275.64$		& $-1022.73$		& $901.92$	\\
P5b 		& $-2470$				& $-407.32$		& $986.75$		& $-1314.84$		& $930.40$		& $2.50$	\\
P5c 		& $-2770$				& $-485.44$		& $1562.38$		& $-2905.32$		& $2883.52$		& $-896.92$	
		\\ \hline \hline
\end{tabular}
\end{center}
\caption{Values of $c_i(\delta)$ with $\delta=1$ fitted to the symmetry energy parameters
$J=32.78$~MeV, $L=49.25$~MeV, $K_{\rm sym}=-156.26$~MeV, $Q_{\rm sym}=583.07$~MeV, and
three $R_{\rm sym}$ values.
The unit of $R_{\rm sym}$ is MeV and $c_i$ is in the unit of $\si{MeV} \cdot \si{fm}^{3+i}$.
}
\label{tab:P5s}
\end{table*}

As a further test, we repeat the process adopted in Sec.~\ref{symmetric}, namely, we now vary the 4th derivative in nuclear symmetry energy, 
the kurtosis $R_{\mathrm{sym}}$ in this section.  
Since the value of $R_{\rm sym}$ obtained from P4 set is about $-2470$~MeV, we consider the variation by $\pm 300$~MeV from this value.
For other parameters, we fix $J=32.78$~MeV, $L=49.25$~MeV, $K_{\rm sym}=-156.26$~MeV, and $Q_{\rm sym}=583.07$~MeV.
Table~\ref{tab:P5s} presents the values of parameters $c_i^{}(1)$ with three different $R_{\rm sym}$ values, which defines P5a, P5b, and P5c. 
For completeness, we use the values of $c_i^{}(0)$ determined as S3b in the previous section.

\begin{figure}[t]
\centering
\includegraphics[width=0.4\textwidth]{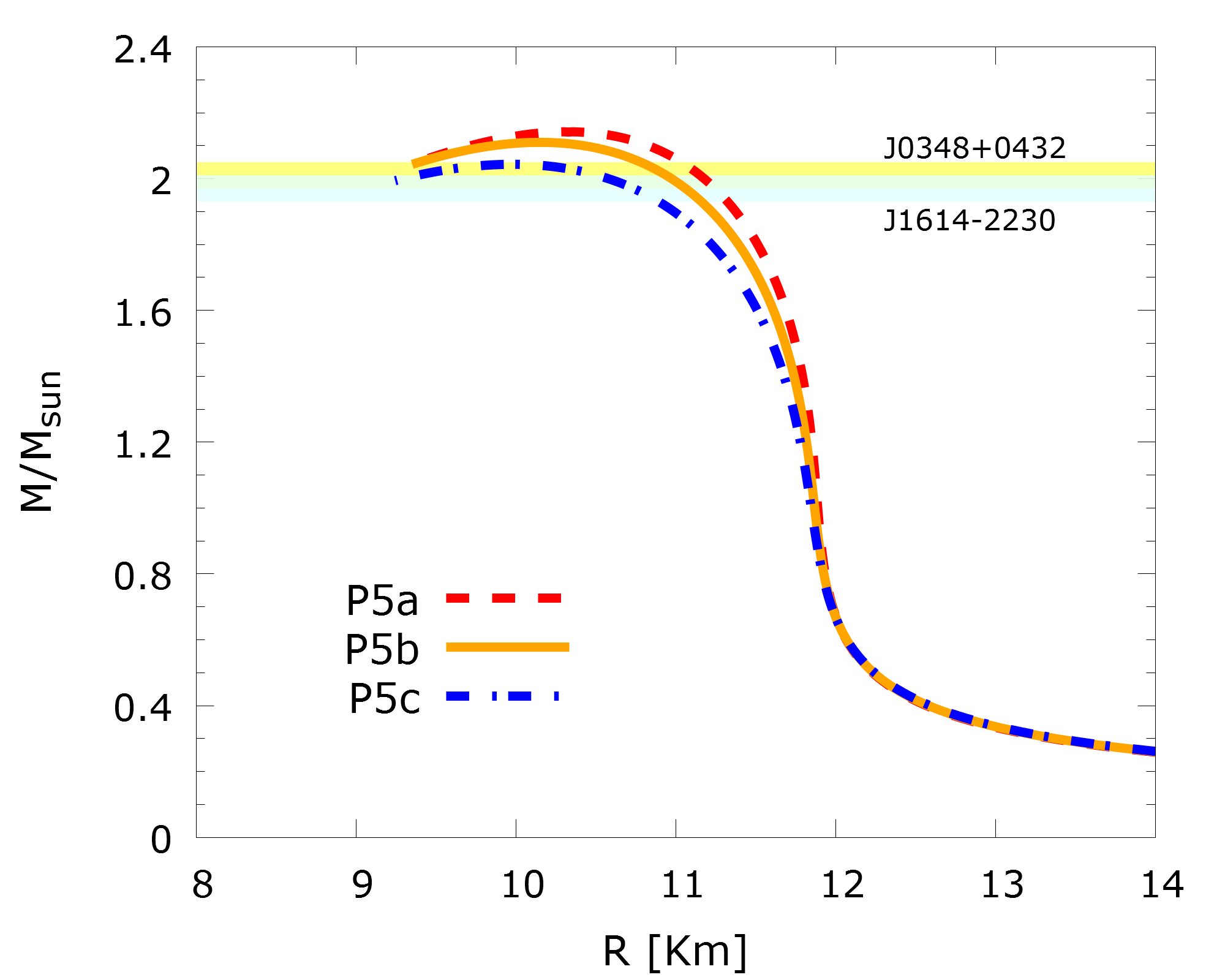}
\caption{Neutron star mass-radius relations: Results corresponding to the parameter sets P5a, P5b, and P5c.}
\label{fig8}
\end{figure}

\begin{figure}[t]
\centering
\includegraphics[width=0.4\textwidth]{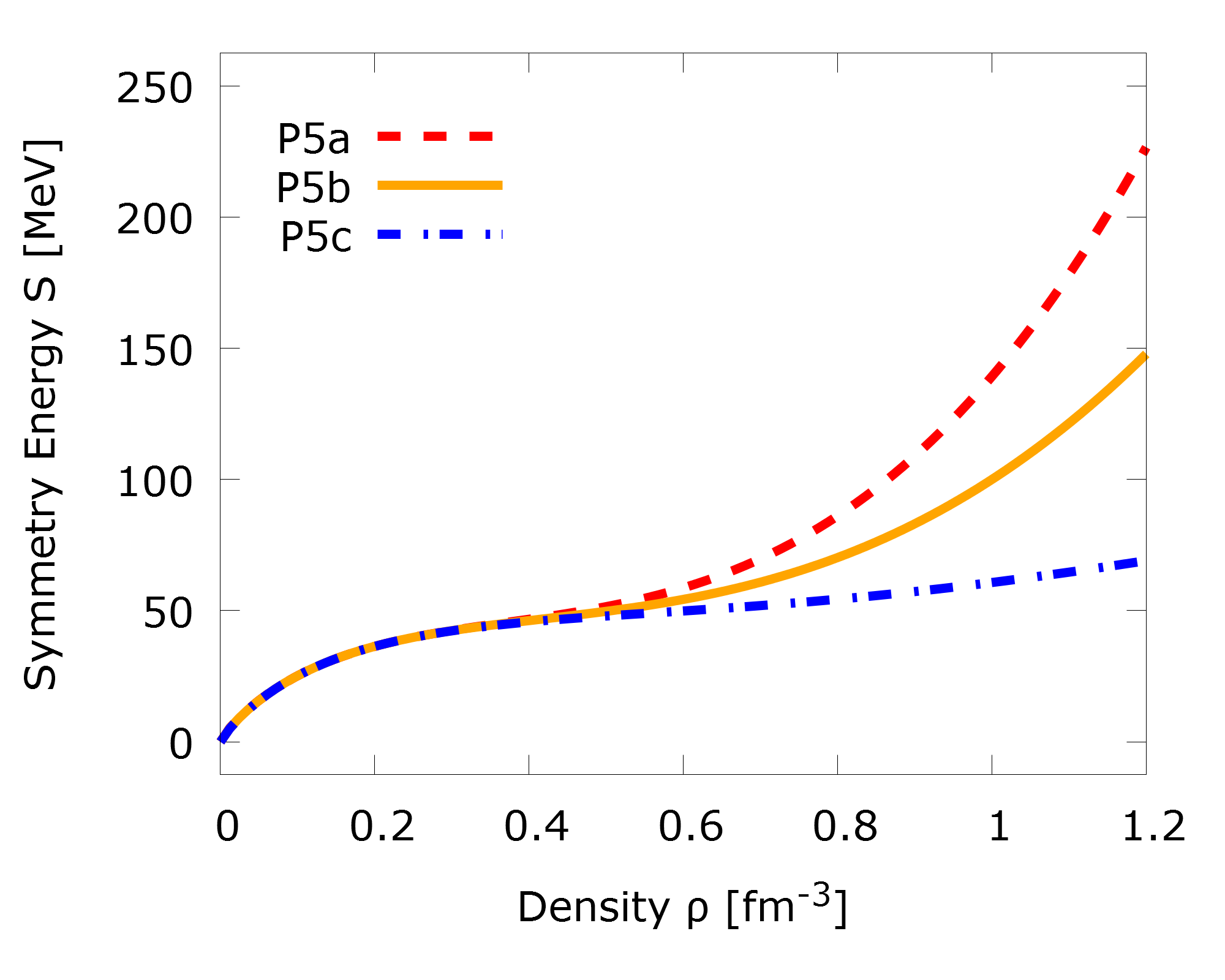}
\caption{Symmetry energy obtained with the parameter sets P5a, P5b, and P5c.}
\label{fig:sym2}
\end{figure}

We first examine the effect of $R_{\rm sym}$ variations on infinite nuclear matter by calculating the neutron-star mass-radius relations.
Figure~\ref{fig8} depicts the predictions on the neutron star mass and radius curves with the models P5a, P5b, and P5c.
All these models predict the maximum mass larger than $2 M_\odot$, which is consistent with the observational constraints
of Refs.~\cite{DPRRH10,AFWT13}.
In order to see the origin of this phenomena, we plot the nuclear symmetry energy for these three models in Fig.~\ref{fig:sym2}.
This clearly shows that varying $R_{\rm sym}$ affects the nuclear symmetry energy only at densities larger than $0.6 \sim
0.7~\mbox{fm}^{-3}$ which is within the range of maximum central density of neutron stars~\cite{LH19}.
Since R5a, R5b, and R5c give similar symmetry energy below this density, they would give similar results for neutron stars.

The effect of varying $R_{\rm sym}$ can also be explored in low-mass neutron stars by considering the tidal deformability.
For a neutron star with a mass of $1.4 M_\odot$, we found that P5a, P5b, and P5c models give the dimensionless tidal deformability of
315.8, 304.1, and 289.4, respectively.
These values are well below the upper limit of the observation, $\Lambda(1.4 M_\odot) \leq 800$, which again originates from the
similarities of symmetry energy of the three models below $0.6 \sim 0.7~\mbox{fm}^{-3}$.

%
\begin{table*}[t]
\renewcommand{\arraystretch}{1.1}
\begin{center}
\begin{tabular}{c|ccc} \hline\hline
Parameter & P5a   & P5b & P5c  \\ \hline
$t_0^{}$ ($\mbox{MeV} \cdot \mbox{fm}^3$) & $-1772.04$		& $-1772.04$	& $-1772.04$ \\
$y_0^{}$ ($\mbox{MeV} \cdot \mbox{fm}^3$) & $-455.27$	&	$-142.77$		& $169.73$  \\
$t_1^{}$ ($\mbox{MeV} \cdot \mbox{fm}^5$) &  $246.82$ & $275.85$	&	$273.68$ \\
$t_2^{}$ ($\mbox{MeV} \cdot \mbox{fm}^5$) & $-173.35$	&	$-161.47$	 & $-162.34$ \\
$t_{31}^{}$ ($10^{4} \, \mbox{MeV} \cdot \mbox{fm}^4$) &  $12216.73$ & $12216.73$ & $12216.73$ \\
$y_{31}^{}$ ($10^{4} \, \mbox{MeV} \cdot \mbox{fm}^4$) &  $2349.81$ & $-11465.31$ & $-25280.43$\\
$t_{32}^{}$ ($ \mbox{MeV} \cdot \mbox{fm}^5$) &  $1099.03$ & $569.01$ & $608.27$ \\
$y_{32}^{}$  ($ 10^{4} \,  \mbox{MeV} \cdot \mbox{fm}^5$) & $-10223.47$	& $28532.42$ & $66660.82$ \\
$y_{33}^{}$ ($10^4 \, \mbox{MeV} \cdot \mbox{fm}^6$) & $24545.47$ & $-22329.53$ & $-69204.53$ \\
$y_{34}^{}$ ($10^4 \, \mbox{MeV} \cdot \mbox{fm}^6$) & $-21646.05$ & $-59.93$ & $21526.20$
\\ \hline
$\zeta$ &  $-0.7116$	& $0.1139$ & $0.0527$ \\
$W_0$ ($\mbox{MeV}\cdot \mbox{fm}^5$) &  $105.57$ & $108.49$ & $107.90$
\\
 \hline \hline
\end{tabular}
\end{center}
\caption{Same as Table~\ref{tab:PNM} but for P5a, P5b, and P5c. Here, $t_{33}^{} = t_{34} = 0$ as we use S3b for $\alpha_i^{} = c_i^{}(0)$}
\label{tab:PNM-2}
\end{table*}

\begin{table*}[t]
\renewcommand{\arraystretch}{1.1}
\begin{center}
\begin{tabular}{c|cccc|cccc} \hline\hline
 \multirow{2}{*}{ Nuclei{\;} }
 			& \multicolumn{4}{c|}{Binding energy per nucleon [MeV]}						
			 & \multicolumn{4}{c}{Charge radius [fm]}								\\
 			\cline{2-9}
 			& Expt. 	& P5a			& P5b			& P5c
			& Expt. 	& P5a			& P5b			& P5c
 			\\ \hline
 \multirow{2}{*}{ $^{40}$Ca }
 			& $8.5513^*$	&  $8.5567$		& $8.5564$		& $8.5564$
			& $3.4776^*$	& $3.4786$		& $3.4781$		& $3.4782$
 			\\
 			&			&  $(0.063\%)$		& $(0.060\%)$		& $(0.060\%)$
			&			& $(0.029\%)$		& $(0.014\%)$		& $(0.018\%)$
			\\ 
 \multirow{2}{*}{ $^{48}$Ca }
			& $8.6667^*$	&  $8.6575$		& $8.6566$		& $8.6560$ 
			& $3.4771^*$	& $3.4872$		& $3.4867$		& $3.4863$
 			\\
 			&			&  $(0.106\%)$		& $(0.117\%)$		& $(0.123\%)$
			&			& $(0.291\%)$		& $(0.276\%)$		& $(0.264\%)$
			\\ 
 \multirow{2}{*}{ $^{208}$Pb }
			& $7.8675^*$	&  $7.8800$		& $7.8806$		& $7.8808$ 
			& $5.5012^*$	& $5.4891$		& $5.4886$		& $5.4880$
 			\\
 			& & $(0.159\%)$		& $(0.167\%)$		& $(0.170\%)$
			&			& $(0.221\%)$		& $(0.229\%)$		& $(0.240\%)$
 			\\
 			\hline\hline
 \multirow{2}{*}{ $^{16}$O }
 			& $7.9762$	&  $7.8633$ 		& $7.8683$		& $7.8679$	
			& $2.6991$	& $2.7636$		& $2.7618$		& $2.7619$
 			\\
 			&			&  $(1.42\%)$		& $(1.35\%)$		& $(1.36\%)$
			&			& $(2.39\%)$		& $(2.32\%)$		& $(2.33\%)$
			\\ 
 $^{28}$O		& ---			& $6.0467$		& $6.0623$		& $6.0746$
 			& ---			& $2.8381$		& $2.8371$		& $2.8353$
 			\\ 
 $^{60}$Ca	& ---			& $7.6470$		& $7.6545$		& $7.6611$ 
 			& ---			& $3.6475$		& $3.6465$		& $3.6451$
			\\ 
 \multirow{2}{*}{ $^{90}$Zr }
			& $8.7100$	& $8.7357$		& $8.7330$		& $8.7322$ 
			& $4.2694$	& $4.2474$		& $4.2476$		& $4.2474$
 			\\
 			&			& $(0.295\%)$		& $(0.265\%)$		& $(0.255\%)$
			&			& $(0.516\%)$		& $(0.511\%)$		& $(0.516\%)$
 			\\ 
 \multirow{2}{*}{ $^{132}$Sn }
  			& $8.3549$	& $8.3539$		& $8.3559$		& $8.3564$ 
			& $4.7093$	& $4.7093$		& $4.7088$		& $4.7082$
 			\\
 			&			& $(0.012\%)$		& $(0.013\%)$		& $(0.019\%)$
			&			& $(0.000\%)$		& $(0.010\%)$		& $(0.024\%)$
			\\ \hline\hline
\end{tabular}
\end{center}
\caption{Same as Table~\ref{tab:nuclei2} but for P5a, P5b, and P5c.
The SNM parameters are fixed to the values of model S3b in Table~\ref{table1}. The experimental data are from Refs.~ \cite{NNDC,AM13}.}
\label{tab:nuclei3}
\end{table*}

\begin{figure}
\includegraphics[width=0.95\columnwidth]{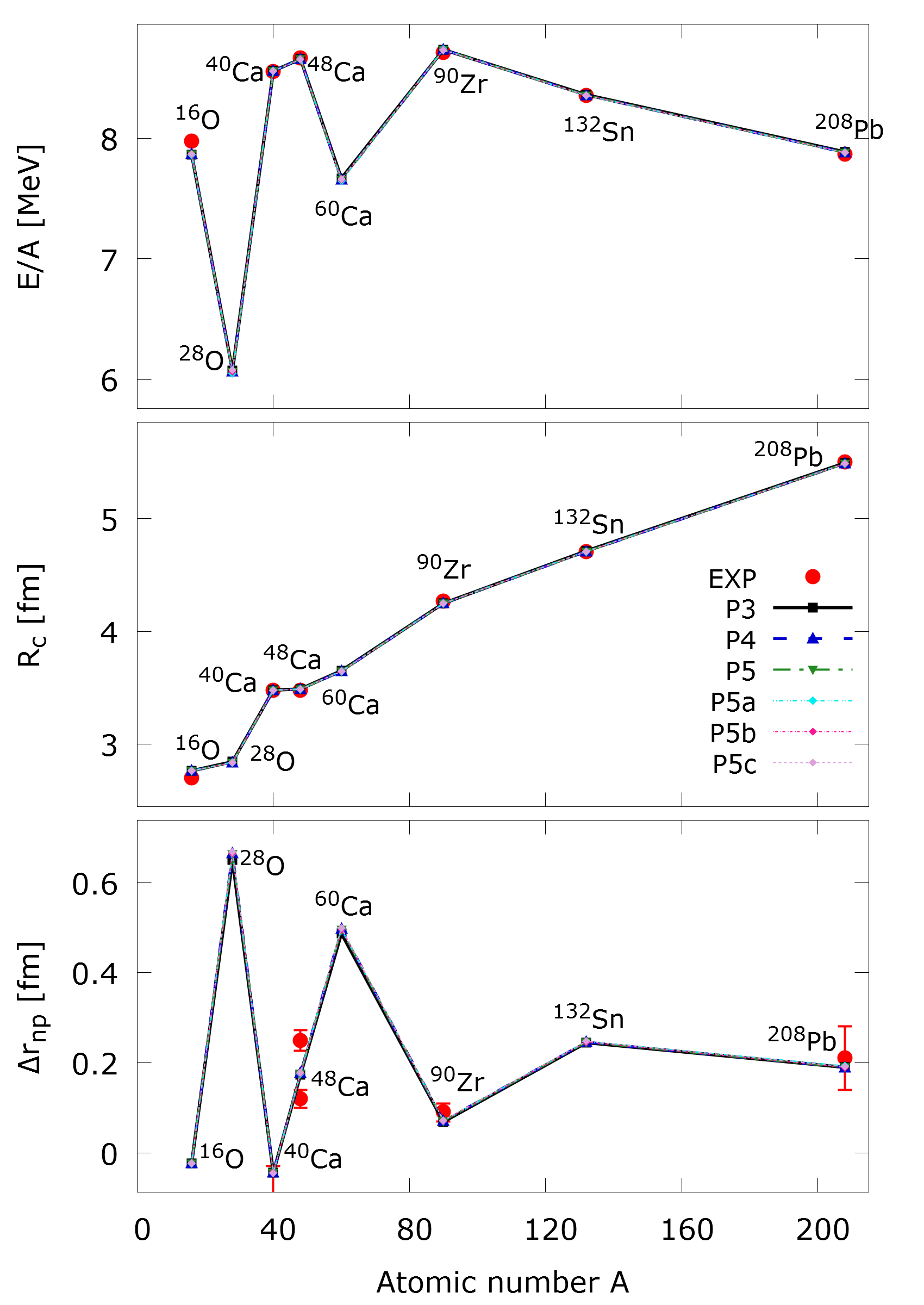}
\caption{Results for $E/A$, $R_c$, and neutron skin thickness $\Delta r_{np}$.
Neutron skin thickness data are from Refs.~\cite{JTLK04,MACD17,PREX12}.
}
\label{fig10}
\end{figure}

Tables~\ref{tab:PNM-2} and \ref{tab:nuclei3} show the fitted parameters and resulting properties of nuclei. 
Here again, we find that the three models give similar results, which leads us to conclude that nuclear properties are quite 
insensitive to $R_{\rm sym}$. 
To illustrate the point visually, we compare in Fig.~\ref{fig10} the neutron skin thickness $\Delta r_{np}$ obtained with 
P3 , P4 (baseline set), P5, P5a, P5b, P5c, together with the results for $E/A$ and $R_c$.
The similarities shown in Fig.~\ref{fig10} imply that the higher order terms in EDF cannot be constrained by normal nuclear data.%
\footnote{We also investigated the dependence of nuclear properties on the value of $R_{\rm sym}$ by allowing more than $\pm 1,000$~MeV
from the value of P5b to confirm that the nuclear properties are not sensitive to $R_{\rm sym}$.}

\section{Discussion\label{discussion}} 

Following the above detailed presentation of results, let us recapitulate what we have done and learned and discuss how our work relates to other current undertakings of similar scope. 

First, we have confirmed that seven EoS parameters suffice for a description of nuclei as well as homogeneous matter in a broad range of densities. 
The number is consistent, on one hand, with the four EoS parameters (plus the surface tension) required in the ``minimal nuclear energy functional"~\cite{BFJPS17} which only concerns finite nuclei; and on the other hand with the conclusions of the recently proposed ``meta-modeling" approach for neutron stars~\cite{MHG17}, namely that the skewness of the EoS plays a non-negligible role, but a less significant one than low-order parameters in the description of neutron stars.   

The analytical form of the KIDS EoS and EDF for homogeneous matter, namely an expansion in powers of the cubic root of the density~\cite{PPLH16}, was inspired by quantum many-body theories and effective field theories. The analytical form allows a straightforward, analytical mapping between the KIDS parameters and an equal number of EoS parameters, see, e.g., Eqs.~(\ref{eq:eosB})$-$(\ref{eq:eosE}). Thus we can vary any of the EoS parameters at will and examine its effect on observables. 
In addition, we are able to vary the effective mass values at will~\cite{GPHO18}, which gives KIDS unprecedented flexibility. 
So far we have applied the KIDS EDF at the Hartree-Fock level for nuclear ground states, but studies of excitations within the random phase approximation are also possible and in progress. 
In this sense our approach goes well beyond the meta-modeling, whose applications in nuclei 
have been limited to semi-classical results for bulk ground-state nuclear properties~\cite{CGRM17,RG17}. 

The description of nuclei was achieved by reverse-engineering a convenient Skyrme-type functional. In the process, the amount of momentum dependence (encoded, for example, in the effective mass value and gradient terms) vs. genuine density dependence (encoding correlations and three-nucleon forces) needs to be determined. Although we have found that static, bulk nuclear properties are 
practically independent of the effective mass~\cite{GPHO18}, the same may not be true for dynamic phenomena such as giant resonances. Studies are in progress~\cite{PG18}. 
Nevertheless, the small amount of momentum relative to density dependence generally favored by our studies so far, undermines the possibility to eliminate density-dependent couplings completely, as is attempted in certain generalizations of the Skyrme functional based on high-order momentum-dependent terms and on the density-matrix expansion~\cite{CDK08,DPN15}. 

Based on our present results we may conclude that a fit of more than the above seven EoS parameters to nuclear data would make little sense. 
(On the contrary, a free fit of all parameters could lead to overfitting.) Although further EoS parameters and a strong momentum dependence are not desired or required, 
 in order to achieve precision, it does make sense to explore extensions of the KIDS EDF for nuclei by including additional effects which are not active (or are weakly active) in homogeneous matter. 
One of them, already included, is the spin-orbit term. Another interesting possibility is the tensor force, as already pursued in modern Skyrme functionals~\cite{LBBDM07}. 
Time-odd terms are also unconstrained at present. Our preferred approach would be to use pseudodata for polarized homogeneous matter. 

\section{Summary\label{summary} and conclusion}

The main purpose of this work was to validate the optimal number of EoS parameters required for a description 
of nuclei and homogeneous matter in a broad range of densities. 
Previous work in the framework of the KIDS EDF had indicated that symmetric nuclear matter could be efficiently modeled with three low-order parameters in an expansion in Fermi momentum and that PNM requires four parameters. 
The conclusion was based solely on a statistical analysis of fits to pseudodata for homogeneous matter. 
In this work, in order to confirm the expansion and its convergence,  
we explored the role of widely used parameters characterizing the EoS at the saturation point. 
In particular, we fixed the saturation density, the energy at saturation and the compression modulus $K_0$ of symmetric matter, as well as the symmetry energy at saturation density $J$, its slope $L$ and  its curvature and skewness, to baseline values and varied the EoS skewness in symmetric matter at saturation, $Q_0$, and the kurtosis of the symmetry energy, $R_{\mathrm{sym}}$. 
We examined the effect in dilute and dense matter (neutron star properties) and on nuclear structure.

In regard to the uncertainty from $Q_0$, its effect is negligible up to $\rho \sim 0.4\, {\rm fm}^{-3}$ ($\sim 2.5 \rho_0$). 
The maximum mass of neutron stars 
shows non-negligible dependence on $Q_0$, but the uncertainty is not significant enough to affect the consistency with existing observations.
No effect on bulk nuclear properties was discerned. 

In the extension of expansion of isospin asymmetric part of EDF, 
the results for $N=6$ showed symptoms of overfitting so we stopped at the fifth term. 
Comparison of $N=3$ fitting result to input data demonstrated that three terms in asymmetric part are insufficient
to guarantee the reproduction of input data but the fits saturate at $N=5$. 
The interpretation is consistent with the EoS of dilute neutron matter, symmetry energy at high densities,
and mass-radius curves of neutron stars.
Again, the choice of kurtosis values $R_{\mathrm{sym}}$ did not affect the description of nuclear properties. 

Bulk properties of spherical magic nuclei were calculated.
Results turned out to be independent of $Q_0$ values,
and the number of terms in asymmetric part of EDF did not affect the prediction for nuclei.
Similar conclusions hold for $R_{\mathrm{sym}}$.

From the present results we conclude that  three terms in the symmetric part, and four terms in
the asymmetric part of the EoS are sufficient for a unified description of both infinite (unpolarized) nuclear
matter and finite nuclei in a single framework. 
Fitting a nuclear EDF with more than the seven necessary EoS parameters to nuclear data 
can arguably lead to overtraining and loss of predictive power. 
The determination of the most realistic values for the minimal EoS parameters can of course be persued with the help of data and statistical analyses.
In addition, extended density dependencies of non-local terms can be explored~\cite{PF94,FPT97,FPT01}.
The EoS of polarized matter is yet another topic to be considered. 
But attempts at refining the nuclear EDF beyond that number of terms must focus on parameters which are not active 
(or strongly active) in unpolarized homogeneous matter, for example the effective tensor and time-odd terms.

\section*{Acknowledgments}
Y.O. is grateful to Yeunhwan Lim for useful information and comments.
This work was supported by the National Research Foundation of Korea under 
Grants Nos.~NRF-2017R1D1A1B03029020, NRF-2018R1D1A1B07048183,
NRF-2018R1A6A1A06024970, and NRF-2018R1A5A1025563. 
The work of P.P. was supported by the Rare Isotope Science Project of the Institute for Basic Science funded by
Ministry of Science and ICT (MSIT) and the National Research Foundation of Korea (No. 2013M7A1A1075764).

\end{document}